\documentstyle[12pt,axodraw,epsfig]{article}
\input epsf
\newcommand{\beq}{\begin{equation}}
\newcommand{\eeq}{\end{equation}}
\def\bea{\begin{eqnarray}}
\def\eea{\end{eqnarray}}
\newcommand{\ba}{\begin{array}}
\newcommand{\ea}{\end{array}}

\def\lsim{\stackrel{<}{{}_\sim}}
\def\gsim{\stackrel{>}{{}_\sim}}

\def\msnu{m_{\tilde \nu}}
\def\mst{m_{\tilde{t}_1}}
\def\msT{m_{\tilde{t}_2}}
\def\msb{m_{\tilde{b}_1}}
\def\msB{m_{\tilde{b}_2}}
\newcommand{\photino}{\tilde{\gamma}}

\begin{document}

\begin{flushright}
{CERN-TH/99-300} \\
{FTUV/99-65}\\
{IFIC/99-68}\\
{hep-ph/9911317}\\

\end{flushright}
\vspace{1cm}
\begin{center}
{\large Neutral Higgs Sector of the MSSM without
$R_p$  }\\
\vspace{.2cm}
\end{center}
\vspace{1cm}
\begin{center}
{Sacha Davidson }\\
{CERN Theory Division,
CH-1211 Gen\`eve 23, Switzerland}\\
\vspace{.5cm}
{Marta Losada}\footnote{On leave of absence from the Universidad Antonio Nari\~{n}o, Santa Fe de Bogot\'a, COLOMBIA.}\\
{CERN Theory Division,
CH-1211 Gen\`eve 23, Switzerland}\\
\vspace{.5cm}
{Nuria Rius }\\
{Depto. de F\'{\i}sica Te\'orica and IFIC, Centro Mixto\\
Universidad de Valencia-CSIC, Valencia, Spain}\\
\vspace{.5cm}

\end{center}
\hspace{3in}

\begin{abstract}
We analyse the neutral scalar sector of the MSSM
without R-parity. Our analysis 
 is performed for a one-generation
model in terms of ``basis-independent'' parameters, and includes
one-loop corrections due to large yukawa couplings. 
We concentrate on the consequences of large $R_p$ violating
masses in the soft sector, which mix the
Higgses with the sleptons, because these are only
constrained by their one-loop contributions to neutrino masses.
We focus on the effect
of  $R_p$-violation on the Higgs mass and branching
ratios. We find that the experimental lower
bound on the lightest CP-even Higgs in this
model can be  lower than in the MSSM.
\end{abstract}

\vspace{1cm}
{November 1999} \\

\newpage

\section{Introduction}

Supersymmetry(SUSY) \cite{Ross, Nilles,HK,Barb} is a popular extension 
of the Standard Model (SM), that introduces
new  scalar partners for  SM fermions and new fermionic 
partners for SM bosons.
A consequence of the enlarged particle content of SUSY
models is that baryon  (B) and lepton (L) number
are not automatically conserved in the
renormalisable Lagrangian. In the  Standard Model,
  gauge invariance implies 
that $B$ and $L$ are conserved in  any terms of dimension
$\leq 4$; this is no longer the case in
SUSY,  so  a discrete symmetry  is often
imposed  to forbid the unwanted interactions
that violate $B$ and/or $L$.

There are a variety of discrete symmetries  \cite{IR} that can be imposed to
remove the renormalisable $B$ and $L$ violating  terms
from the SUSY  Lagrangian.
The most common is $R$-parity \cite{fayet}, under which particles  have
the charge $R_p \equiv (-1)^{3B+L+2S}$, where $S$ is the spin. 
SM particles are even under this transformation, and
SUSY partners are odd, which forces SUSY particles  to
always be made in pairs and forbids the Lightest
Supersymmetric Particle (LSP) from decaying.

Alternatively, one can allow the $B$ and $L$
violating interactions to remain in the SUSY Lagrangian,
and constrain  the couplings to be consistent
with present experimental data. The renormalisable
$R_p$ violating couplings violate either $B$, or $L$.
If both types of coupling are simultaneously present, they
can mediate proton decay, and are therefore
constrained to be very small \cite{Smirnov}. 
So in this paper, we will assume that the
$B$ violating couplings are absent---forbidden
by some other symmetry--- and only consider
the $L$ violating couplings. These are particularly interesting,
because $L$ violation is observed in neutrino masses.

The  renormalisable $R_p$ violating
interactions have a variety of phenomenological
consequences \cite{hall}. These include
generating majorana neutrino masses, 
mediating various flavour and
lepton number violating processes \cite{herbi,bhat,french},
and modifying the signatures of supersymmetric particles
at colliders \cite{french,NillesNir}
In particular it allows
the lightest supersymmetric particle (LSP)
to decay \cite{BGH,LSP}. It can also modify the Higgs
sector.

The Higgs sector of the $R_p$-conserving 
MSSM has been extensively studied 
\cite{HK,haber,dawson,hempfling,espinosa,Carena1,bagger,okada}, 
with a lot of emphasis on both one-loop 
\cite{brignole,ellis,zwirner1,chankowski,Carena1} and more
recently on two-loop effects 
\cite{espinosa, hempfling2,hempfling3,Casasetal,Carena2,hollik,Carena3} 
to the lightest Higgs boson mass. 
The most relevant one-loop effects due to the large top-quark 
Yukawa coupling are from
the stop-top sector. There are several different approaches 
that have been utilised to
incorporate these loop effects: effective potential methods, 
renormalisation group running, explicit diagrammatic
calculations (see {\it e.g.} \cite{carenazerwas} for
a review). The effective potential, which we
use here,  can in a simple way 
take into account the most relevant effects
although it does not incorporate any momentum-dependent 
contributions.

A Higgs boson could be the
next particle discovered at accelerators. It is
therefore interesting to study its properties in various
extensions of the Standard Model, in particular SUSY.
One of the advantages of 
the supersymmetric Standard Model for cosmology is that baryogenesis
may be possible at the electroweak phase transition---
if the Higgs  is light enough \cite{dfl,BAU}.  However,
as the experimental lower limit  on the Higgs mass  
increases, the parameter space remaining in the MSSM
for baryogenesis is reduced. Adding
$R_p$ violation can decrease the experimental lower limit
on the Higgs mass, which could increase the available parameter
space for electroweak baryogenesis.

In this paper, we study the neutral Higgs sector
at one-loop in the $R_p$-violating MSSM with one generation
of quarks and leptons, 
since this toy model already contains the main 
effects of the complete three generation case. 
We vary bilinear and trilinear  $R_p$ violating parameters
over their experimentally allowed ranges, and discuss
how this can change the masses 
of the neutral CP-even scalar bosons 
and the branching ratios of the lightest one, $h_1$. 
The $R_p$-violating Higgs sector has been studied
by numerous authors: 
novel decays of both neutral \cite{fer,drw}
and charged \cite{andy} scalar bosons have been 
analysed in the context of bi-linear $R_p$ violation,  
and the mass matrices of the Higgs sector  
have been derived, considering only the effect of 
bi-linear terms \cite{CF1} and in the general case 
with both bi-and tri-linear couplings \cite{CF2}.
Our analysis differs from  previous
treatments in that we include one-loop yukawa corrections 
to the Higgs masses, and we parametrise $R_p$
violation in a basis-independent way that avoids 
much possible confusion about what is a lepton/slepton
in a lepton number non-conserving theory.

The next section of this  paper introduces our notation and
discusses the basis-independent approach to
$R_p$ violation. The third section
is devoted to experimental constraints on
the $R_p$ violating parameters in our model,
largely from neutrino masses. 
In the fourth and fifth sections, we calculate the
masses and various branching ratios for
the CP-even Higgses at one loop. We present our
results in section six. The first appendix
contains the one-loop Higgs mass matrices
in an arbitrary basis.
The second appendix contains the same
information, but in the basis where the sneutrino vev is zero (to
one loop).
The third appendix contains
a few useful but long formulae.

\section{Basis dependence of the Lagrangian }

In the SM, the Higgs and leptons have the same gauge quantum numbers.
However, they cannot mix  because the Higgs is 
a boson and the leptons are fermions. In a supersymmetric model
this distinction is removed, so the down-type
Higgs and sleptons can be assembled in a vector $L_J = (H_d, L_i)$ 
with $J:0..N_g= $ the number of generations.
We write vectors in $L_J$ space with a capitalised index $J$ or
as vectors $\vec{v}$, and we write matrices in $L_J$ space
in bold face {\bf m}.
Using this notation,
the superpotential for the  supersymmetric
SM with   $R_p$ violation  can be written as
\beq
W= \mu^J {H}_u  L_J + \lambda_{\tau}^{JK \ell} L_J L_K E^c_{\ell} + 
\lambda_b^{Jpq} L_JQ_p D^c_q  + h_t^{pq} {H}_u Q_p U^c_q \label{S}
\eeq
The $R_p$ violating and
conserving coupling constants
have been assembled into vectors and matrices in $L_J$ space:
we call the usual $\mu$ parameter $\mu_0$,
and identify the usual $\epsilon_i =  \mu_i$, 
$\frac{1}{2}h_e^{jk} = \lambda_{\tau}^{0jk}$, $\lambda^{ijk} = 
\lambda_{\tau}^{ijk}, $  $h_d^{pq} = \lambda_b^{0pq}$, 
 and $\lambda^{'ipq} = \lambda_b^{ipq}$.
Lower case roman indices $i,j,k$ and $p,q$  are lepton and
quark generation indices. In the body of the paper,
we will work in a one generation model, so
$\frac{1}{2}h_{\tau} = \lambda_{\tau}^{01}$, 
  $h_b = \lambda_b^{0}$, 
 and $\lambda^{'} = \lambda_b^{1}$
and now the capitalised indices run
from 0..1, and 1 corresponds to the third lepton
generation.
 We often write $d$ and $L$
(for down-type Higgs and slepton) rather
than 0 and 1. $Q$, $U^c$ and $D^c$ are
the third generation quark superfields.
 In the one-generation  model, there is 
no $\lambda LLE^c$ interaction
(because $\lambda$ is antisymmetric on the capitalised
indices).

We also include possible $R_p$ violating couplings among
the soft SUSY breaking parameters, which can
be written as
\beq
V_{soft} = \frac{\tilde{m}_u^2}{2} H_u^{\dagger} H_u + \frac{1}{2}
 L^{J \dagger} [\tilde{m}^2_L]_{JK} L^K  + B^J H_u L_J 
 + A_t H_u Q U^c + 
     A_b^{J} L_J Q D^c +
    A_{\tau}^{JK} L_J L_K E^c + h.c.  \label{soft}
\eeq
Note that we have absorbed the superpotential parameters
 into the $A$ and $B$ terms; $e.g.$ we write $B^0 H_uH_d$
not $B^0 \mu^0 H_u H_d$ ~~\footnote{We do this because $B_J$
is a vector---a one index object---in $\{L_J\}$ space.
>From this perspective, giving it two indices can lead to confusion.}.
We abusively use capitals for superfields (as in (\ref{S})) and
for their scalar components. 

The reason we have put the Higgs $H_d$ into a vector with the sleptons,
and combined the $R_p$ -violating with the $R_p$ conserving
couplings, is that the lepton
number violation can be moved around the Lagrangian
by judiciously choosing which linear combination
of  hypercharge = -1  doublets  to identify as the
Higgs/higgsino, with the remaining doublets
being sleptons/leptons. This can create
some confusion when one tries to set experimental
constraints on lepton number violating couplings; 
it makes little sense to set an upper bound on
a coupling constant that can be made  zero by a basis
rotation.

If one calculates a physical observable as a function
of measurable quantities, then the basis in which
one does the intermediate steps of the calculation
is irrelevant. However, if one computes
observables as a function of Lagrangian quantities,
as is common in Supersymmetry, it can be  important to specify 
the basis  chosen in the Lagrangian. 
In SUSY theories with lepton number violation,
there are various possible choices for what
one identifies as a lepton/slepton in the Lagrangian,
and the interactions that are ``lepton number violating''
depend on this identification. However, this freedom
to redefine what violates $L$ 
is deceptive,  because phenomenologically we know that
the leptons are the mass eigenstate $e, \mu$ and $\tau$ ,
so we know what lepton number violation is.
There are two possible approaches to this fictitious
freedom; either one chooses to work
in a Lagrangian basis that corresponds to the
mass eigenstate basis of the leptons,
or one can construct combinations
of coupling constants that are independent
of the basis choice to parametrise the $R_p$
violation in the Lagrangian \cite{NillesNir,Banks,DE1,DE2,D3}. 
These invariant
measures of $R_p$ violation in the Lagrangian
are analogous  to Jarlskog invariants  
which parametrise  CP violation.

 The standard option  is to
work in a basis that corresponds approximately
to the mass eigenstate basis of the leptons. 
For instance, if one chooses the Higgs direction in
$L_J$ space to be parallel to $\mu_J$, then the 
additional bilinears in the superpotential $\mu_i$
will be zero. In this basis, the sneutrino vevs are
constrained to be small by the
neutrino masses, so this is approximately
the lepton mass eigenstate basis. Lepton number violation
among the fermion tree-level masses
in this basis is small
by construction, so it makes sense to neglect
the bilinear $R_p$ violation, or treat 
the small $R_p$ violating masses as ``interactions''
within perturbation theory, and set constraints on  the trilinears,
as is commonly done (for a review, see {\it.e.g.} \cite{herbi,bhat}.
For a careful analysis including the bilinears, see \cite{BKMO}).

In this paper, we present 
our results in terms of basis-independent
``invariants''. We also give explicit results in the
basis where the sneutrino does not have a vev, 
which is  close to the lepton mass eigenstate
basis.
This is to present  our calculation in a familiar way. 
 The advantage of the first approach
is that we can express Higgs masses and branching ratios
in terms of inputs that are independent of
the choice of basis in the Lagrangian. The drawback is that the 
``invariants'' can appear unwieldy
and forbiddingly complicated. However, since we
work in a model with only one lepton generation,
the linear algebra is tractable. 

The aim of the ``basis-independent'' approach is
to construct
combinations of coupling constants that
are invariant under rotations in
$L_I$ space, in terms of which one can express
physical observables. By judiciously combining
coupling constants one can find ``invariants'' which
are zero if $R_p$ is conserved,  so these invariants parametrise
$R_p$ violation in a basis-independent way.
For instance,  consider the superpotential of equation (\ref{S})
in the one generation limit, $I:0..1$.
 It appears
to have two $R_p$ violating interactions:
$\mu_{1} H_u L$ and $\lambda' LQD^c$.  It is
well known that one of these can be rotated into
the other by mixing $H_d$ and $L$  \cite{hall}. If
\bea
H_d' = \frac{\mu_0}{\sqrt{\mu_0^2 + \mu_{1}^2}} H_d +  
        \frac{\mu_{1}}{\sqrt{\mu_0^2 + \mu_{1}^2}} L \nonumber \\
L' = \frac{\mu_{1}}{\sqrt{\mu_0^2 + \mu_{1}^2}} H_d -  \label{simple}
        \frac{\mu_0}{\sqrt{\mu_0^2 + \mu_{1}^2}} L ~~~,
\eea
then the Lagrangian expressed in terms of $H_d'$ and
$L'$ contains no $  H_u L'$ term. One could instead dispose 
 of the $\lambda' L QD^c$ term.
The coupling constant
combination that is invariant under basis redefinitions
in $(H_d,~L)$ space, zero if $R$ parity is conserved,
and non-zero if it is not is $\mu_0 \lambda' - h_d \mu_{1} =
( \mu_0, \mu_{1}) \wedge ( h_d, \lambda')$.

In this paper, we are interested in $R_p$ violating effects
in the Higgs sector, so we are interested in constructing
invariants involving $B_J$, the $L_J$  mass matrix 
   [{\bf m}$_L^2]_{JK} \equiv [\tilde{m}_L^2]_{JK} + \mu_J \mu_K$, 
and the $L_J$ vev $v_J \equiv \langle L_J^0 \rangle$.
The  vev $v_J$ is a dependent
variable, fixed by  $B_J$ and  [{\bf m}$_L^2]_{JK}$
in the minimisation conditions.
 In an arbitrary basis,
there are therefore two $R_p$ violating masses
in the Higgs sector: $B_1$ and  [{\bf m}$_L^2]_{01}$.
However one can always choose the basis such that one of these 
parameters is zero, so we
expect only one independent  invariant parametrising $R_p$ violation
in the (tree-level) Higgs mass matrices.

\begin{figure}
\begin{picture}(400,300)(-150,-100)
%the two vectors
\thicklines
\put(0,0){\vector(3,4){100}}
\put(0,0){\vector(0,3){150}}
%the axes
\thinlines
\put(0,-50){\line(0,3){250}}
\put(-50,0){\line(3,0){250}}
%the projection lines for the vectors
\put(-30,120){$v$}
\put(40,85){$B$}
\put(5,20){$\eta$}
\put(10,180){$H_d$}
\put(170,10){$L$}
\end{picture}
\caption{Non-orthogonality of the soft mass $B_J H_u L^J $ and
the $H_d$-slepton vev $\langle L_J \rangle = v_J$. 
The angle $\eta$ between $\vec{B}$ and $\vec{v}$ is  
basis-independent. The ``invariant'' 
$ R = v^2 B^2 - (\vec{v}\cdot \vec{B})^2$  is equal to
$v^2 B^2 \sin^2 \eta$.
The basis here is $ \hat{H} \propto \vec{v}$, 
and  $ \hat{L} \propto \vec{v} \cdot \lambda_{\tau}$.}
\label{f1}
\end{figure}
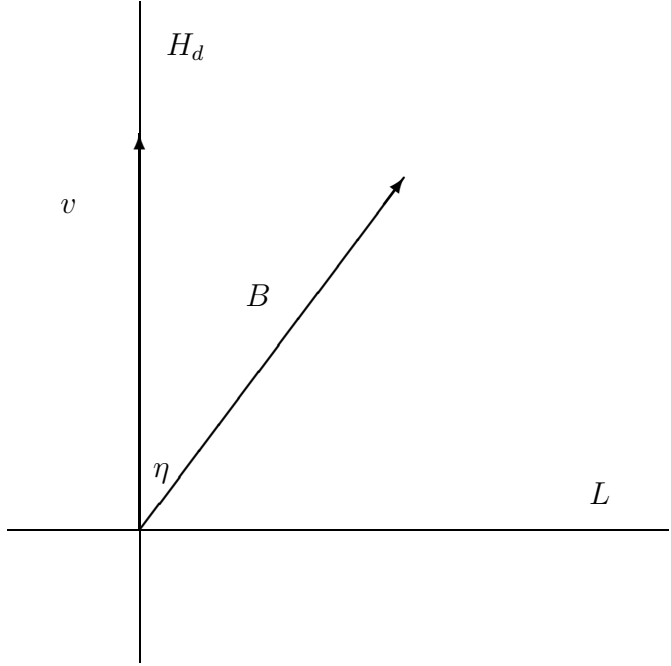

There is $R_p$ violation in the Higgs  sector if 
$\vec{B}$, [{\bf m}$_L^2]$, 
and $\vec{v}$  disagree on which direction
in $L_J$ space is the Higgs, or equivalently, if it  
is not possible to choose a basis 
where $v_L = B_L = $[{\bf m}$^2]_{dL} =0$.
$\vec{B}$ is a vector that would like to be the Higgs---that 
is, if the basis in $L_I$ space
is chosen such that $H_d \propto \vec{B}$ then $B_d = |B|$ and
$B_L = 0$, so the mass matrix mixes $H_u$ with 
$H_d$ but not with $L$. [{\bf m}$_L^2]_{JK}$ has two eigenvectors
in $L_J$ space, one of which would like to be the Higgs, 
and the other the slepton. $ \vec{v}$ is 
also a candidate direction in $L_J$ space
to be the Higgs---the basis where $H_d$ is the
$\vec{v}$ direction is the basis where the sleptons do not
have vevs. There is $R_p$ violation if  two of $\vec{B}$, $\vec{v}$ and 
[{\bf m}$_L^2]_{JK}$ do not agree on what is the Higgs direction.
A  convenient choice for the invariant parametrising
this $R_p$ violation at tree-level is 
\beq
R = v^2 |\vec{B}|^2 - (\vec{v}\cdot \vec{B})^2 ~~~; 
\delta_R = \frac{R}{ v^2 B^2} ~~~,
\label{R1}
\eeq
where $\delta_R$ is the normalised version of the
parameter, varying from 0 for no $R_p$ violation to
1 for maximal $R_p$ violation.  
As we will see  from the minimisation conditions 
(equations \ref{mincdn1} and \ref{mincdn2}), at tree level
$\chi \vec{B} = - [${\bf m}$_L^2] \cdot \vec{v}$, 
where $\chi$ is the vev of the up-type Higgs, 
so we can also write
$\chi^2 R = v^2 \vec{v} \cdot  [$ {\bf m}$_L^2]^2 \cdot \vec{v} - (\vec{v} 
\cdot [${\bf m}$_L^2] \cdot \vec{v})^2$.
$R$ parametrises the $R_p$ violation in the 
 mass matrix  relevant for the Higgs. $\sqrt{\delta_R}$ is  the
sine of the angle \footnote{We take the positive square root:
$\sin \eta = + \sqrt{\delta_R}$.}
between $\vec{B}$ and $\vec{v}$ 
(see figure \ref{f1}), which is
clearly independent of the choice of basis in
$L_J$ space.

 There are many other invariants that 
parametrise $R_p$ violation
among other coupling
constants. For instance, there is an additional
invariant among the bilinears in one generation
 \cite{Banks,NillesNir}. There are
three possible directions in $L_J$ space that could be identified
as the Higgs:
$B_J, \mu_J$ and one of the eigenvectors of [{\bf m}$_L^2]_{JK}$. 
If these three vectors  do not coincide,
there should be two invariants parametrising 
the misalignment between the three vectors. 
One in the scalar sector, as constructed in equation
(\ref{R1}), and an additional one involving $\vec{\mu}$.
 For instance, if $\vec{\mu}$ is misaligned 
with respect to $\vec{v}$, mixing between
neutrinos and neutralinos generates a tree-level neutrino mass
$\sim \vec{\mu} \wedge \vec{v} = $
$ v \cdot \lambda_\tau \cdot \mu/|\lambda_\tau|$. Invariants 
 parametrising $R_p$ violation between bilinears and trilinears can
also be constructed.  Since
the upper bound on neutrino masses constrains $
 \vec{\mu} \wedge \vec{v}  $ to be small, we neglect it in this paper,
and concentrate on the effects of $\delta_R$.

Up to this point, we have discussed the construction
of invariants using parameters from the Lagrangian
without specifying whether they were tree-level,
or computed to some loop order.  We choose
to  write the invariants in terms of
one-loop parameters. We do this because 
the invariants were constructed to avoid
 expressing measurable quantities ({\it e.g.} masses)
 in terms of
unmeasurable  basis dependent  Lagrangian parameters. 
So we  define the invariants in terms of
one-loop parameters, because these are closer to
what is physically measured.  The invariants $R$
and $\delta_R$ discussed above are therefore
taken to be
\beq
 R = v^2 \vec{M}_u^2 - (\vec{v}\cdot  \vec{M}_u)^2 ~~~; 
\delta_R = \frac{R}{ v^2  \vec{M}_u^2} ~~~,
\label{R}
\eeq
where $\vec{M}_u$ is the one-loop corrected
version of $\vec{B}$ that appears in the CP-odd mass matrix
(\ref{CPodd}). From the one-loop minimisation conditions (\ref{mincdn1})
and (\ref{mincdn2}),
$\vec{M}_u = - ${\bf M}$\cdot \vec{v}/\chi$,
where {\bf M} is the one-loop version of {\bf m}$_L^2$
that appears in the CP-odd mass matrix. $R$ can therefore
also be written as
\beq
\chi^2 R = v^2 \vec{v} \cdot  {\bf M}^2 \cdot \vec{v} - (\vec{v} 
\cdot {\bf M} \cdot \vec{v})^2.
\eeq
The drawback to
using  the one-loop expressions is that  
it is not obvious which loop corrections
should be included. 

We will use $\delta_R$ rather than $R$ as our $R_p$ violating
parameter, because it is dimensionless and normalised to 1.
For small $\delta_R$, this is clearly a good choice, because
the magnitude of $\vec{M}_u$ is largely determined
by its $R_p$ conserving component ($\sim m_A \sin \beta \cos \beta$
 in the MSSM).
However, as $\delta_R $ increases to 1,  the magnitude 
of the $R_p$ violating  mass$^2$  term  $|\vec{M}_u| \sqrt{\delta_R}$ can
nonetheless decrease if $|\vec{M}_u|$ does.   We will see
that for some parameter choices,  this is the case.

We would like to determine which are the necessary conditions
on the $R_p$ violating parameters to produce  a substantial effect
on physical observables.
Hence,  we do not assume in this paper that $B_I  \approx B \mu_I$,
(and $ [\tilde{m}_L^2] \approx \tilde{m}^2$ {\bf I})
 as would be expected in  many  models of SUSY
breaking.  This means that we allow $\delta_R$ to be as
large as experimentally allowed.

\section{Experimental constraints}

Both low and high-energy processes  can place stringent
bounds (see $e.g.$ \cite{herbi,bhat})
 on the $R_p$-violating couplings which give rise to new interactions.
The most relevant constraints on the $R_p$ violating
bilinear couplings come from neutrino masses. The trilinear
$\lambda'$ also contributes to neutrino masses,
but the most restrictive bound on $\lambda'$ comes
from $Z$ decay to $b \bar{b}$. We now
mention the contribution to
neutrino masses due to various
$R_p$ violating parameters; the purpose of
this discussion is to set bounds on our
parameters, not to calculate the neutrino mass.

In $R_p$-violating models the neutrino can acquire a mass at tree-level
through mixing with the neutralinos  and
also through loops which violate lepton number by two units.
In the basis where the sneutrino vevs are zero, 
the tree-level contribution can be written as \cite{Banks, Nardi,Nowakowski}
\beq
m_{\nu_{\tau}} = {m_Z^2 \mu_{0} M_{\photino}\cos^2\beta\over
m_Z^2M_{\photino} \sin 2\beta - M_{1}M_{2} \mu_{0}} \sin^2 \xi
\label{treemnu}
\eeq
where $M_1$, $M_2$ are gaugino masses, 
$M_{\photino} = M_1 \cos^2\theta_{W} + M_2 \sin^2 \theta_{W}$, 
and $\sin \xi = (\vec{\mu} \wedge \vec{v})/|\mu||v|$.
Thus the neutrino mass sets the constraint
that $\vec{\mu}$ be aligned with $\vec{v}$, which determines the
tree-level contribution, without imposing any constraints
on the other $R_p$-violating parameters.

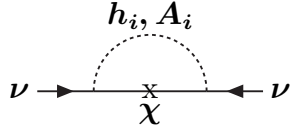
\begin{figure}[ht]
\unitlength.5mm
\SetScale{1.418}
\begin{boldmath}
\begin{center}
\begin{picture}(60,40)(0,0)
\ArrowLine(0,0)(15,0)
\Line(45,0)(15,0)
\ArrowLine(60,0)(45,0)
\DashCArc(30,0)(15,0,180){1}
\Text(-2,0)[r]{$\nu$}
\Text(62,0)[l]{$\nu$}
\Text(30,20)[c]{$h_i, A_i$}
\Text(30,0)[c]{x}
\Text(30,-6)[c]{$\chi$}
\end{picture}
\end{center}
\end{boldmath}
\vspace{10mm}
\caption{Contribution to the one-loop neutrino mass from bilinear 
$R_p$ violation in the soft masses. This diagram is possible
because the sneutrinos mix with the
Higgses. }
\label{figmnu}
\end{figure}

There is also a loop contribution to the
neutrino mass proportional to $\delta_R$,
as discussed in \cite{GH1,GH2,KK}. 
If $\vec{v}$ and $\vec{M}_u$
are not parallel, then
the $R_p$ violation in the soft masses will mix the
real (imaginary)  part  of the sneutrino with the CP-even (odd)
Higgses. This introduces a mass splitting between 
$\tilde{\nu}_R = Re(\tilde{\nu})$ and 
$\tilde{\nu}_I = Im(\tilde{\nu})$. A neutrino
mass can be generated by a neutralino-neutral scalar
loop ---see figure (\ref{figmnu}). The amplitude for this diagram is
\beq
m_{\nu} = \frac{g^2}{64 \pi^2} \sum_{\chi_j} m_{\chi_j}
(Z_{j2} - Z_{j1} g'/g)^2
\sum_i(\hat{\nu} \cdot \hat{s}_i)^2  \epsilon_i B_0(0, M^2_i,m^2_{\chi_j})
\label{numassexact}
\eeq
We in practise neglect the sum over the
four neutralinos and just include the
lightest one. 
The $Z_{ij}$ are the usual mixing angles between the neutralino
mass and interaction eigenstate bases---for
simplicity we only include the
gauge coupling of the neutralino.
We sum over three CP-even and two CP-odd scalars : 
$s_i = \{ h_1, h_2, h_3,A_1, A_2\}$.
The $\{ (\hat{\nu} \cdot \hat{s}_i ) \}$ are
the mixing angles between the neutrino
and the various scalars $ s_i$. They
are basis-independent quantities which we will
calculate as dot products
in $L_J$ space in  section 5.
$\epsilon_i$ is $+1$ for the three CP-even
Higgses and $-1$ for the CP-odd.
$B_0$ is  a Passarino-Veltman function:
\beq
 B_0(0, M_s^2,m^2_{\chi}) 
= - 16 \pi^2 i \lim_{q \rightarrow 0} \;
\int \frac{ d^{2\omega} k}{(2 \pi)^{2 \omega}}
\frac{1}{[(k+q)^2 - m_{\chi}^2](k^2 - M_s^2)}
\supset  - \frac{M_s^2}{M_s^2 - m_{\chi}^2} 
\ln \left( \frac{M_s^2}{ m_{\chi}^2} \right) ~~~.
\label{B0}
\eeq
There are  divergent and scale-dependent
contributions to $B_0$ in addition to the
right hand side of equation (\ref{B0}); however
these cancel in the sum  over scalars
$s_i$ in equation (\ref{numassexact}).

The dependence of $m_{\nu}$ on different parameters can
be understood in various limits. As $\delta_R \rightarrow 0$
two of the CP even neutral scalars, $h_i$ and $h_j$, 
become $h$ and $H$ of the MSSM,
the third CP even scalar $h_k$ becomes $\tilde{\nu}_R$, and
$A_1 \rightarrow A$ of the MSSM while 
 $A_2 \rightarrow \tilde{\nu}_I$. The overlap
between the neutrino and the MSSM Higgs
$\{h_i,h_j,A_1 \}$ goes to zero (we will show in the 
next section that it is $\propto \delta_R$),
and  $\hat{\nu} \cdot \tilde{\nu}_R \sim \hat{\nu} \cdot \tilde{\nu}_I$
$\rightarrow 1$. The real and imaginary parts of the sneutrino contribute
to the sum with opposite sign; we expect $m^2_{h_k} -
m^2_{A_2} \propto \delta_R$ so the
sneutrino contribution
will also go to zero with $\delta_R$ \cite{GH1,GH2}.

The neutrino mass also decreases (for arbitrary $\delta_R$)
as either the neutralino mass or the CP-odd scalar mass $m_{A_2}$ goes
to infinity. We can therefore estimate
\beq
m_{\nu} \lsim \frac{g^2  m_{\chi} \delta_R}{64 \pi^2}  \left\{
\begin{array}{cc}
      \frac{m_{Z}^2}{m^2_{A_2}} & m_{A_2} 
          \rightarrow \infty \\
             & \\
      \frac{m^2_{Z}}{m_{\chi}^2} & {m_{\chi}} 
          \rightarrow \infty 
\end{array}  \right.
\eeq
where we have assumed that as $m_{A_2} $ or $m_{\chi}$
become large, all  remaining masses are of order $m_Z$.
This is an overestimate, because it neglects cancellations
in the sum (\ref{numassexact}). 
For $\delta_R \sim 1$, and most choices of $m_{A_1},
m_{A_2}, m_{\chi}$ and $\tan \beta$, the neutrino
mass will be $< 10 $ MeV, so there is
no bound on $\delta_R$
from the laboratory limit $m_{\nu_{\tau}} < 10$ MeV.
If we require $m_{\nu} \lsim $ eV, as would be
required by oscillation data, we find $\delta_R \lsim
10^{-6}$ for  $m_{A_2} \sim m_{\chi} \sim m_Z$.

The second type of loop diagrams involve fermion-sfermion loops.
The contribution  proportional 
to the trilinear coupling 
constant $\lambda'_{i33}$ in the
usual three-generation mass-eigenstate
basis-analysis,  
can be expressed in a basis-invariant way as
\beq
m_{\nu_{\tau}}^{loop} =  {3\over 16 \pi^2} X_{b} 
{f(x) \over \msB^2} \; (\hat{\nu}\cdot\vec{\lambda})^2 m_{b} \ , 
\label{mnut1}
\eeq 
where $f(x) = - {\log x\over 1-x}$, $x= \biggl({\msb\over \msB}\biggr)^2$,
$X_b$ is given in appendix A, and $\hat{\nu}$ and $\vec{\lambda}$ are 
defined in section 5. In the basis where the sneutrino does not have a vev, 
$\hat{\nu}= (0,-1)$ and $\vec{\lambda}=(h_b,\lambda')$. Here $\hat{\nu}$ is 
specifying the neutrino direction.
So, $\lambda'_{i33}$ will have a certain allowed upper value for a given
set of the inputs that determine the sbottom mass parameters. 

Another bound on $\lambda'_{i33}$ has been given in the literature
 from the calculation of  $R_{l}$ \cite{BES} in 
which the allowed value of the coupling
scales with right-handed soft-SUSY breaking mass 
$m_{B_R}$. For 
  $m_{B_R} \gsim 500$ GeV
the trilinear coupling can be of order 1. Other bounds
from $B_{o}$--$\bar{B}_{o}$ mixing or 
$B\rightarrow \tau \bar{\nu} X$ 
\cite{Agashe,Feng,Ligeti} 
also have been studied and they also allow values of
 $\lambda'_{i33} \sim 1$ for sufficiently 
heavy right-handed sbottom, on the order of 300 GeV
\footnote{The  bounds from references \cite{Agashe,Feng,Ligeti} 
depend on whether the CKM-mixing is present in 
the down-quark sector.} .

The actual numerical bounds on the $R_p$ violating coupling 
will depend on the input value one takes for 
the neutrino mass. If we use the
experimental limit on the tau neutrino 
mass $\sim 10$ MeV, 
we can easily have a value $\vec{\lambda} \cdot \hat{\nu} \sim 1$, 
thus its effect on the Higgs sector
will be analogous to that of the top Yukawa 
coupling. In this case the bound from $R_l$ 
is stronger for generic values of the $R$-parity 
conserving parameters. For smaller values of
the neutrino mass, such that for example 
neutrino oscillation scenarios can be 
fulfilled, the bounds are very strong on 
the $R_p$ violating couplings.

Note that allowing $\lambda' \sim 1$ in the 
fermion mass eigenstate basis means that $\vec{\lambda}$ 
is almost perpendicular to $\vec{v}$. The
$b$-quark mass is 
$m_b = - (\vec{\lambda} \cdot \vec{v})/\sqrt{2} \ll |\vec{v}|$,
and $\lambda' = \vec{v} \wedge \vec{\lambda}/|\vec{v}|$.

There are various accelerator limits on
particle masses and coupling constants when
$R$-parity is not conserved 
(see $e.g.$ \cite{french} for a discussion).
These often depend sensitively on a number
of parameters, so are difficult to translate
to the model we consider here. We will
comment the LEP lower bound on the mass of sneutrinos
with  $R_p$ violating decays in section 6.

\section{Higgs boson masses}

In this section, we calculate the
Higgs boson  masses  using the effective potential.
To do this
we make an $SU(2)$ rotation on the $H_d$ doublet,
$ H_d \rightarrow \Phi_d = \varepsilon H_d^* ~ (\varepsilon_{12} = -1,$ 
$\varepsilon^2 = -1)$, 
to put the neutral component in the same element of
the doublet as for $H_u$. This makes it easy to
compare the $R_p$ conserving part of
our calculation to standard two-Higgs doublet results. 
 We also rotate the   slepton field. So 
we can write
\beq
H_u = \left( \begin{array}{c} H_u^+ \\ H_u^0 \end{array} \right) = 
 \left( \begin{array}{c} \Phi_u^+ \\ (\chi + \phi_u^R + i \phi^I_u)/{\sqrt{2}}
 \end{array} \right),~~~~~~~~
H_d = \left( \begin{array}{c} H_d^0 \\ H_d^- \end{array} \right) = 
 \left( \begin{array}{c}  ( v_d +\phi_d^R -i \phi^I_d)/{\sqrt{2}}
\\ - (\Phi_d^+)^* \end{array} \right) 
\label{rotn}
\eeq
\beq
L = \left( \begin{array}{c} L^0 \\  L^- \end{array} \right) = 
 \left( \begin{array}{c}  (v_L + \phi_L^R - i \phi^I_L)/{\sqrt{2}} 
\\ - (\Phi_L^+)^* \end{array} \right)
\eeq
where we define $\chi$  to be
the $H_u$ vev, and $v_d, v_L$ to be the down-type Higgs
and slepton vevs (in some arbitrary basis). We will not be concerned
with the charged fields in this paper.

From the superpotential and soft terms of equations
(\ref{S}) and (\ref{soft}), the tree-level potential
for the neutral  scalar vevs is
\beq
V_{tree} = m_u^2 \frac{\chi^2}{2}   + \frac{1}{2} \vec{v} \cdot 
[{\bf m}^2_L] \cdot \vec{v} 
+ \chi \vec{B} \cdot \vec{v} + \frac{\Lambda}{4} (\chi^2 - v^2)^2 
\label{Vtree}
\eeq
where  $\Lambda = (g^2 + g^{'2})/8$
,$v^2 = |\vec{v}|^2= v_d^2 + v_L^2$,
$m_u^2 = \tilde{m}^2_u + |\vec{\mu}^2|$  and 
[{\bf m}$^2_L]_{JK} = [\tilde{m}^2_L]_{JK} + \mu_J \mu_K$.

We include the loop corrections due to
large yukawa-type
couplings, but not due to gauge couplings. The one-loop
contribution to the potential from tops, stops,
bottoms and sbottoms will be
\begin{eqnarray}
V_{loop}&  =& \frac{1}{64 \pi^2} 
\left( -12 m_t^4 \left[ \ln \left( \frac{m_t^2}{Q^2} \right) \right.
- \frac{3}{2} \right]   +6 m_{\tilde{t}_1}^4 \left[ \ln \left( 
\frac{m_{\tilde{t}_1}^2}{Q^2}\right)- \frac{3}{2} \right] +6 m_{\tilde{t}_2}^4 \left[ \ln \left( 
\frac{m_{\tilde{t}_2}^2}{Q^2}\right)- \frac{3}{2} \right] \nonumber \\ & &
-12 m_b^4 \left[ \ln \left( \frac{m_b^2}{Q^2} \right) 
- \frac{3}{2} \right]  \left. +6 m_{\tilde{b}_1}^4 \left[ \ln \left( 
\frac{m_{\tilde{b}_1}^2}{Q^2} \right) 
- \frac{3}{2} \right] +6 m_{\tilde{b}_2}^4 \left[ \ln \left( 
\frac{m_{\tilde{b}_2}^2}{Q^2} \right) 
- \frac{3}{2} \right] \right)~~~~. \label{Vloop}
\end{eqnarray}
We include the bottom contributions because the $R_p$ violating
$\lambda'$ can be large (in the basis where the sneutrino
does not have a vev).

We are principally interested in the  behaviour of
the lightest CP-even neutral scalar---the ``Higgs''. 
We would like to obtain its mass as a function of observables
 like the masses of the CP-odd scalars, and parameters
like $\tan \beta$ and   the ``invariant'' $\delta_R$ (equation 
\ref{R}) that parametrises
$R_p$ violation. We therefore need the $3 \times 3$
mass matrices for the CP-even and CP-odd Higgses at
the minimum of the potential.

The tree-level minimisation conditions can be
written in terms of  the CP-odd mass
matrix elements (\ref{CPodd}).
In the absence of CP violation,
the one-loop minimisation conditions expressed in terms of
the one-loop CP-odd mass matrix have the same functional form
(see equations (\ref{mincdn1}) and (\ref{mincdn2})).
This is useful because it means we can impose
the minimisation conditions at one loop
without calculating either the one-loop
CP-odd mass matrix or the one-loop minimisation conditions.
To see this,  we write the potential
as a function of six variables:
\beq
C_1 = H_u^{0*} H_u^{0}, C_2=H_d^{0*} H_d^{0},C_3= L^{0*} L^{0},
 C_4 = H_u^{0} H_d^{0}, C_5=H_u^{0} L^{0},C_6= H_d^{0*} L^{0} ~~~.
\eeq
The three minimisation conditions for  the potential
can then be written
\begin{eqnarray}
0 & = & {\partial  V \over \partial H_u^0 }\equiv \sum_{n=1}^6 \frac{\partial V}{\partial C_n} \frac{\partial C_n}{\partial \chi} \\
0 & =&  {\partial  V \over \partial L_J^0 }\equiv \sum_{n=1}^6 \frac{\partial V}{\partial C_n} \frac{\partial C_n}{\partial v^J}~~~J= 0,1.
\end{eqnarray}

The CP-odd mass matrix is of the form
\beq
M^{\mathrm{CP-odd}}=\left[
\begin{array}{cc}
 M_{uu} & \left( \begin{array}{cc} M_{ud} & M_{uL} 
           \end{array} \right) \\
 \left( \begin{array}{c} M_{ud} \\ M_{uL}  \end{array} \right) & 
            \left( \begin{array}{cc} {\bf M}_{dd} & {\bf M}_{dL} \\ 
              {\bf M}_{dL} & {\bf M}_{LL} \end{array} \right)
\end{array}
\right]
\label{CPodd}
\eeq
where the individual elements are
\beq
M_{ij} =  \frac{\partial^2 V}{\partial \phi^I_i \partial \phi^I_j} 
   =  \sum_{n=1}^6 \frac{\partial V}{\partial C_n} 
          \frac{\partial^2 C_n}{\partial \phi^I_i \partial \phi^I_j } 
\eeq
Note that our capitalised $M$s have  mass dimension  2.
The indices $i,j$ run from 1..3, or over 
$u,d,L$, and the $\{\phi^I_i \}$
are the imaginary parts of the scalars (see equation \ref{rotn});
``$I$'' is not an index in $L_J$ space.)
Second derivatives of $V$ do not appear because they
are multiplied by first derivatives of the $\{ C_i \}$,
which are zero (evaluated at $\phi^I_j = 0$). 
Since 
\beq
\frac{\partial C_1}{\partial \chi} = \chi \frac{\partial^2 C_1}
{\partial \phi^I_u \partial \phi^I_u }
\eeq
(and similarly for  the other derivatives of the $\{ C_n\}$), 
we see that the minimisation conditions can be written
in terms of the CP-odd mass matrix:
\beq
 M_{uu} +  \frac{M_{u J}v^J}{\chi} = 0 
\label{mincdn1}
\eeq
\beq
 M_{uJ} +  \frac{{\bf M}_{JK}v^K}{\chi} = 0 ~~~~.
\label{mincdn2}
\eeq
We emphasize  that these equations are valid in any basis,
and we apply them at one-loop.

Explicit formulae for the minimisation 
conditions and the mass matrix elements can be
found in the Appendices. Appendix A 
contains the results for an arbitrary basis in
terms of basis-invariant quantities. 
In appendix B we present the results in the basis $v_L=0$, 
using the familiar Lagrangian notation. 

The eigenvalues of the CP-odd mass matrix $M$ are easy
to obtain, since $M$ has a zero eigenvalue.
The two non-zero eigenvalues are
\beq
m^2_{A_1}, m_{A_2}^2 =
 \frac{1}{2} \left[ M_{uu} + Tr[{\bf M}] \pm \sqrt{ (M_{uu} + Tr[{\bf M}])^2
-4 (M_{uu} Tr[{\bf M}]  + det[ { \bf M }]  -   |\vec{M}_u|^2) }\right]~~~,
\label{mAmsnu}
\eeq
In the $R_p$ conserving limit, $m_{A_2} \equiv m_{\tilde{\nu}}$.
When $R_p$ is not conserved, the sneutrino as a complex field
has Dirac and Majorana masses, so its real and imaginary parts
are not degenerate. The mass of the imaginary part is
what we identify here as $m_{A_2}$.
By using the minimisation conditions (\ref{mincdn1}) and 
(\ref{mincdn2}),  we can rewrite these masses 
in terms of ``basis-independent''
invariants (scalars in $L_J$ space) as
\beq
m^2_{A_1}, m_{A_2}^2 = \frac{1}{2} 
\left( \frac{\vec{v} \cdot [{\bf M}] \cdot \vec{v}}{\chi^2}
+  Tr[{\bf M}] \pm \sqrt{\left( \frac{\vec{v} \cdot {\bf M} \cdot \vec{v}}{\chi^2} 
\frac{2 \chi^2 + v^2}{v^2} - Tr[ {\bf M} ] \right)^2 + \frac{4 R}
{  v^2 \cos^2 \beta}} \right)
\label{21}
\eeq  

Note that we have chosen to write $m_{A_1}^2$ and $m_{A_2}^2$ as
functions of scalars
in $L_J$ space which are non-zero in an $R_p$ conserving theory
(such as $Tr[${\bf M}], $\vec{v} \cdot$ {\bf M}$\cdot \vec{v}$)
and scalars that are zero in an $R_p$-conserving theory ($\delta_R$).
This is slightly different from choosing a basis in which one writes
the masses as a part depending on $R_p$ conserving
couplings and a part
depending on $R_p$ violating couplings (as done
for instance in \cite{Ferrandis}), because for some basis choices the
$R_p$ conserving invariants depend on $R_p$ violating couplings 
($e.g.$ in the $M_{uL} = 0$ basis, $\vec{v} \cdot ${\bf M}$\cdot \vec{v} =
v_d^2$ {\bf M}$_{dd} + 2 v_d v_L$ {\bf M}$_{dL}+v_L^2$ {\bf M}$_{LL}$).

The CP-even mass matrix  will be
\beq
M^{'}_{ij} = M_{ij} + \sum_{n=1}^6 
 \frac{\partial C_n}{\partial Y_i}\times \left(
\frac{\partial }{\partial Y_j} \frac{\partial V}{\partial C_n} \right)
\label{CPeven1}
\eeq
where we have temporarily introduced $Y_i = ( \chi, v_d, v_L)$.
Explicit formulae can be found in the Appendices.
We can express the eigenvalues of the CP-even mass
matrix $ = \{ m_{h_1}^2, m_{h_2}^2, m_{h_3}^2 \}$ in terms
of scalars in $L_J$ space, 
by constructing the characteristic equation of 
({\bf M}$^{'} - m^2$ {\bf I}), and expressing
the coefficients in terms of
invariants.  We do not show
the formulae (analogous to (\ref{21})),
 because they are  too long to be enlightening.
Another possible  way to solve for 
 $\{ m_{h_i}^2 \}$ as a function of
$m_{A_1}^2, m_{A_2}^2, \tan \beta, \delta_R$ and loop corrections,
is to  express the matrix elements
of $M^{'}$ in a basis-invariant way
using equation (\ref{CPeven1}).
We plot the CP-even masses for various inputs
in section 6.

We have chosen $m_{A_1}^2, m_{A_2}^2, \tan \beta, $ 
and $\delta_R$ as inputs because they are ``physical''. 
However there are relations between these parameters
which  constrain the ranges over which they
can be varied. To solve for $m_{h_i}$ as a
function of our inputs,  
we  invert  equation (\ref{21}) to write the
basis invariant $ S \equiv \vec{v} \cdot \vec{M}_u$  
in terms of $m_{A_1}$, $m_{A_2}$ and $\delta_R$:
\beq
S = \frac{\vec{v} \cdot [{\bf M}] \cdot \vec{v} }{\chi^2} 
= \frac{\cos^2 \beta}{2(1 + \gamma)} 
{ \left(m_{A_1}^2 + m_{A_2}^2 
\pm \sqrt{ (m_{A_1}^2 -m_{A_2}^2)^2 
- 4 m_{A_1}^2 m_{A_2}^2  \gamma} \right) }
\label{vMv}
\eeq 
where $\gamma = {\sin^2 \beta} 
{\delta_R / (1-\delta_{R})}$, and the + (-) sign corresponds to 
$m_{A_1}^2 > m_{A_2}^2$ ($m_{A_1}^2 < m_{A_2}^2$).  
$S \rightarrow m_{A_1}^2 \cos^2 \beta$
when $\delta_R \rightarrow 0$.
Clearly this inner product must be a real number;
to ensure that the square root is positive, we need
\beq
\frac{|m_{A_1}^2 - m_{A_2}^2|}{2 m_{A_1} m_{A_2} \sin \beta} > 
\sqrt{\frac{ \delta_R}{1-\delta_R}}
\label{reAbd}
\eeq 
so  $m_{A_1}$ and  $m_{A_2}$ cannot be degenerate
for non-zero $\delta_R$.

\section{Higgs Branching Ratios}

Including $R_p$ violation in the Higgs sector will
modify the interactions  as well as the masses of the Higgses.
Intuitively, it mixes the sneutrino with the neutral Higgses,
so it can modify the amplitudes for Higgs production and
for $R_p$ conserving decays, as well as allowing new
 decay modes such as $h\rightarrow \nu \chi^0$
and $h \rightarrow \tau \chi^+$ \cite{fer,drw}. 
$R_p$ violating couplings also
modify the decays of Higgs decay products. For instance,
the LSP $\chi_0$, produced
in $h \rightarrow \chi^0 \nu$ and  
$h \rightarrow \chi^0 \chi^0$ , could decay
(to three fermions)
within the detector \cite{BGH,LSP}.
It turns into  a neutrino and an off-shell
$h_i$, which then decays to   SM fermions.
So if  $\chi^0 $ can be produced via
an $R_p$ violating vertex (in our case related to $\delta_R$), then 
it decays rapidly through the same
vertex.

The Higgs production and decay rates  clearly
 cannot depend on the basis  in which  they are
 computed, so we will work in a ``basis-independent''
approach. We are principally interested in $R_p$ violation
from the scalar Higgs sector, as parametrised by
the invariant $\delta_R$ of equation \ref{R}, so we will
write the decay rates in terms of this and
other invariants.  There are three  mass eigenstate
bases in $(H_u, L_J)$ space  that are 
relevant for   calculating branching ratios:
the CP-even mass eigenstate basis, the CP-odd basis,
and the fermion mass eigenstate basis. Rotation
angles between these bases will appear
in the Higgs interaction vertices. We will provide
expressions for these  (``physical'') angles 
which are independent of the basis choice in the Lagrangian.

In the $R_p$-conserving MSSM, the lightest CP-even Higgs $h$
is a linear combination of the up and down type
neutral Higgses: $h = \cos \alpha \phi_u^R - \sin \alpha \phi_d^R.$
The $ZZh$ vertex via which LEP can  produce
a $Z$ and an $h$ is
\beq
\frac{ig^2}{2 \cos^2 \theta_W} (\chi \cos \alpha  - v \sin \alpha) 
= \frac{i g m_Z}{ \cos \theta_W}
(\sin \beta \cos \alpha - \cos \beta \sin \alpha)
\label{MSSMZZh}
\eeq
Single sneutrinos cannot be produced in
the MSSM, but the $Z$ can decay to a pair of them
if kinematically possible.
The $Z$ can similarly decay into a CP-even and odd
Higgs, for which the vertex is proportional to $ g \cos 
( \beta - \alpha)/(2 \cos  \theta_W)$.

Adding $R_p$ violation involving one lepton generation
means the sneutrino mixes with the Higgses, so the
lightest Higgs $h_1$ will be a linear combination of
three fields  $\hat{h}_1 = (\cos \alpha) \phi_u^R$ $ - 
(\sin \alpha \cos \varphi) \phi_d^R $ $
 - (\sin \alpha \sin \varphi)  \phi_L^R$. If we define the
angle $\varphi$ with respect to the basis in
$L_J$ space where the sneutrino does not have a vev,
then  $\sin \alpha \cos \varphi = - \hat{h}_1 \cdot \vec{v}/|\vec{v}| $
and $ \sin \alpha \sin \varphi =  - \hat{h}_1 \cdot \hat{\nu} $.
The vector $\hat{\nu}$ is the lepton direction orthogonal
to the vev: $\hat{\nu} = \varepsilon^T \cdot \vec{v}/|\vec{v}|$.
%= \lambda_{\tau} \cdot \vec{v}/| \lambda_{\tau} \cdot \vec{v}\;| $.
If  $ \vec{v} \wedge \vec{\mu} = 0$,
this vector corresponds to the charged lepton mass eigenstate
\cite{DE1,DE2}, which is the neutrino flavour eigenstate.
We therefore call this direction $\hat{\nu}$.
  The $ZZh_1$ vertex is then
a simple generalisation of (\ref{MSSMZZh}):
\beq
\frac{ig m_Z}{ \cos \theta_W} (\sin \beta \cos \alpha  +\cos \beta
  \frac{\vec{v}}{v} 
 \cdot  \hat{h}_1) 
\label{RpZZh}
\eeq
and the $Zh_1A_1$ vertex becomes
\beq
\frac{g}{2 \cos \theta_W} (  \cos \beta \cos \alpha  
% + \hat{\nu}  \cdot  \hat{h}_1) ( p_A - p_h)^{\mu} 
 - (\hat{v}  \cdot  \hat{h}_1) (\hat{v}  \cdot  \hat{A}_1)
 - (\hat{\nu}  \cdot  \hat{h}_1) (\hat{\nu}  \cdot  \hat{A}_1)) 
( p_A - p_h)^{\mu} 
\label{RpZAh}
\eeq
where $p_h$ and $p_A$ are the momenta of the
outgoing scalars.

To evaluate the angles between the CP-even
mass eigenstate basis and the  zero-sneutrino-vev
basis, we must identify the direction in 
$L_J$ space corresponding to $h_1$.
 The lightest eigenvector of
the CP-even Higgs mass matrix  satisfies
\beq
\left[
\begin{array}{cc}
M'_{uu} & \left( \begin{array}{cc} M'_{ud} & M'_{uL} 
           \end{array} \right) \\
 \left( \begin{array}{c} M'_{ud} \\ M'_{uL}  \end{array} \right) & 
            \left( \begin{array}{cc} {\bf M}'_{dd} & {\bf M}'_{dL} \\ 
               {\bf M}'_{dL} & {\bf M}'_{LL} \end{array} \right)
\end{array}
\right] 
\left(\begin{array}{c} u_1 \\ {h}_{1d}  \\ {h}_{1L}
\end{array} \right) = m_{h_1}^2 \left(\begin{array}{c} u_1 \\ {h}_{1d} \\ 
{h}_{1L}
\end{array} \right)
\label{CPeven}
\eeq
where the mass matrix has primes to denote that
it is the CP-even mass matrix and not the CP-odd
matrix of equation \ref{CPodd}. We would like to solve this for
$\vec{h}_1 = (h_{1d}, h_{1L})$ \footnote{ Normalised
vectors wear hats, so for instance $|\hat{h}_1|^2 = 1$.
The mass eigenvectors $\hat{h}_i, \hat{A}_j$ are in
the 3-d ($H_u, L_J$) space; $\vec{h}_i$ is the
projection on $L_J$ space and $|\vec{h}_1|^2 = \sin^2 \alpha$}.
We can write this as two equations for scalars,
vectors, and matrices in $L_J$ space:
\beq
M^{'}_{uu} u_1 +  \vec{M}^{'}_{u} \cdot \vec{h}_1 = m_{h_1}^2 u_1
\label{h2}
\eeq
and 
\beq 
u_1  \vec{M}^{'}_{u} + {\bf M}^{'} \cdot \vec{h}_1 = m_{h_1}^2 \vec{h}_1~~~~.
\label{h3}
\eeq
Rearranging  (\ref{h3}), we find
\beq
\vec{h}_1 =u_1  [m_{h_1}^2 {\bf I} -  {\bf M}^{'}]^{-1} \cdot  \vec{M}^{'}_{u}
\label{hhat}
\eeq
In a one generation model, this is simple
to solve because the inverse of a symmetric $2 \times 2$
matrix {\bf N}$^{'} \equiv [m_{h_1}^2${\bf I} $-$ {\bf M}$^{'}]$
 is {\bf N}$^{'-1} = 
- \varepsilon $ {\bf N}$^{'} \varepsilon/$det({\bf N}$^{'}$),
where $\varepsilon_{11} = \varepsilon_{22} = 0, 
 \varepsilon_{12} = -\varepsilon_{21} = -1$.
So 
\beq
\hat{\nu} \cdot \hat{h}_1 = \frac{u_1}{v \; det[{\bf N}^{'}]} \;
\vec{v} \cdot {\bf N}^{'} \cdot \varepsilon \cdot \vec{M}_u^{'} \ .
\label{nudoth}
\eeq 
and 
\beq
\frac{\vec{v}}{v} \cdot \hat{h}_1 = \frac{u_1}{v \; det[{\bf N}^{'}]} \; 
\vec{v} \cdot   \varepsilon^T 
\cdot {\bf N}^{'} \cdot \varepsilon \cdot \vec{M}_u^{'} \ .
\label{vdoth1}
\eeq 

$u_1 = \cos \alpha$ can be determined from the normalisation of $h_1$:
$u_1^2 + \vec{h}_1^2 = 1$. The vector $\vec{M}_u^{'}
= \vec{M}_u - m_Z^2 \cos \beta \sin \beta ~ \vec{v}/v +$ loop corrections, and
 {\bf M}$^{'}_{IJ} = $  {\bf M}$_{IJ} + m_Z^2 \cos^2 \beta 
\frac{v_I v_J}{v^2}$ 
+ loop corrections, where
$ \vec{M}_u$ and  {\bf M} are from the CP-odd
mass matrix (\ref{CPodd}). These loop corrections, which
are not presented in our analytic formulae, are listed
in the Appendix.  The loop contribution
to the CP-odd mass matrix is implicitly included;
the contribution missing from our
analytic formulae is the  one-loop difference between the
CP-even and CP-odd mass matrices.
Using the minimisation conditions
(\ref{mincdn1}) and (\ref{mincdn2}), we find
\beq
\hat{\nu} \cdot \hat{h}_1  = \frac{u_1}{ det[{\bf N}^{'}] }
S \tan \beta (m_{h_1}^2 + m_Z^2 (\sin^2 \beta -  \cos^2 \beta)) \sqrt{\frac{\delta_R}{ 1 - \delta_R}} + {\rm loop~corrections}
\label{nudoth2}
\eeq
and
\begin{eqnarray}
\frac{ \vec{v} \cdot \hat{h}_1}{v}& = &
\frac{ u_1}{det[{\bf N}^{'}]}
[(m_{A_1}^2 m_{A_2}^2 \sin \beta \cos \beta - m_{h_1}^2 
(S \tan \beta  
 + m_Z^2 \cos \beta \sin \beta ) \nonumber \\ & &  + m_Z^2 \sin \beta 
\cos \beta (m_{A_1}^2 + m_{A_2}^2 - S/\cos^2 \beta )] 
 + {\rm loops}.
\label{vdoth2}
\end{eqnarray}
We do not present formulae for the loop corrections, but they
are included in our numerical plots.
$S$ is defined in equation (\ref{vMv}); the normalisation
factor $u_1/det$[{\bf N}$^{'}]$ is in Appendix C.

In the limit $\delta_R \rightarrow 0$,
the lightest CP-even Higgs $h_1$
can become either the MSSM Higgs $h$ or
the real component of the sneutrino $\tilde{\nu}_R$.
Suppose first that $h_1 \rightarrow h$
as  $\delta_R \rightarrow 0$. Then
 as expected  $\hat{\nu} \cdot \hat{h}_1 \propto 
\delta_R$.
If $h_1 \rightarrow \tilde{\nu}_R$ 
in the $\delta_R \rightarrow 0$ limit,
then $\vec{v} \cdot \hat{h}_1 \rightarrow 0$
because $m_{h_1} \rightarrow m_{A_2}$. 
$\hat{\nu} \cdot \hat{h}_1 \rightarrow 1$ in the same limit,
although this is less obvious because
$u_1/det[${\bf N}$^{'}$] is singular.

 To calculate the contribution of the various Higgses
to the neutrino mass, as discussed in the experimental bounds section,
we need the angle mixing the neutrino with each of the Higgses:
$\hat{\nu} \cdot \hat{s}_i$ ($s_i = \{ h_1, h_2, h_3, A_1,A_2 \})$. 
 These can be computed in the same
way as  $\hat{\nu} \cdot \hat{h}_1$. For $h_2$ and $h_3$,
the formulae are the same,  substituting $m_{h_2}$ or
$m_{h_3}$ for $m_{h_1}$.  For $A_1$ and $A_2$, {\bf M}$^{'}$ is
replaced by {\bf M} and $\vec{M}_u^{'}$ by  $\vec{M}_u$
in the analogue of equation  (\ref{hhat}). This gives
\beq
\hat{\nu} \cdot \hat{A}_i = 
 n_i  S  m_{A_i}^2 \tan \beta    
\sqrt{\frac{\delta_R}{1 - \delta_R}}
+ loops
\label{nudotA}
\eeq
where the normalisation factor $n_i$ is in Appendix C.

There is a technical catch to this way of calculating
the $\{ \hat{\nu} \cdot \hat{s}_i \}$ in the $\delta_R 
\rightarrow 0$ limit. If $\delta_R = 0$,
one of the $h_i$, say $h_3$, and $A_2$
are the sneutrino so  $\hat{\nu} \cdot \hat{A}_2 = \hat{\nu} \cdot \hat{h}_3
= 1$. This is the $\delta_R \rightarrow 0$ limit
of equations  (\ref{nudotA}) and (\ref{nudoth})
because the denominator $\rightarrow 0$, 
but  at $\delta_R = 0$ the equations are singular. 
This can be avoided
by taking $\hat{\nu} \cdot \hat{A}_2 = \sqrt{ 1 - 
(\hat{\nu} \cdot \hat{A}_1)^2}$ and  $\hat{\nu} \cdot \hat{h}_3 = \sqrt{ 1 - 
(\hat{\nu} \cdot \hat{h}_1)^2 - (\hat{\nu} \cdot \hat{h}_2)^2}$
which follow from the unitarity of the rotation matrix.

The tree-level rate for  a scalar $h$ to
decay to two fermions $f_1$ and $f_2$  through a 
vertex of the form
\beq
 h \bar{f}_1( \lambda_L P_L +  \lambda_R P_R)  f_2  ~~~,
\label{vertex}
\eeq
where $P_L = (1-\gamma_5)/2$,  is
\beq
\Gamma (h \rightarrow \bar{f}_1 f_2) =  \frac{1}{8 \pi m_h^2} 
\sqrt{E_2^2 - m_2^2} \left[ (m_h^2 - m_2^2 - m_1^2) 
(\lambda_L^2 + \lambda_R^2) - 4 \lambda_L \lambda_R m_1 m_2 \right]
\label{Gamma}
\eeq
where $E_2 = (m_h^2 + m_2^2 - m_1^2)/(2 m_h)$.

The $R_p$ violating decay rates
$h \rightarrow   \nu \chi^0,  \tau \chi^+$
are both detectable if kinematically allowed, 
because   $\chi^0$  can decay to  $ \nu$ and
an off-shell Higgs, which can then decay to
SM fermions. Here we mention again that 
the neutralino/chargino is produced and decays 
via the same vertex  which
is proportional to $\delta_R$.
If $\delta_R \neq 0$ but
$\vec{\mu} \wedge \vec{v} = 0$ \footnote{$\vec{\mu} \wedge \vec{v} = 0$ 
implies there is no $R_p$ violation at tree level in the
-ino mass matrices.},
the decays $h \rightarrow   \nu \chi^0,  \tau \chi^+$
proceed because the mass eigenstate $h$ 
contains a ``(s)neutrino component'' $ = \hat{\nu} \cdot \hat{h}$.
The coupling constant for the vertex $h \bar{\chi}^0 \nu$ 
is therefore 
\beq
\lambda_L = \lambda_R =\frac{g}{2} (Z_{12} - Z_{11} g'/g)  
\ \hat{\nu} \cdot \hat{h} ~~~~,
\eeq
where $Z$ diagonalises the neutralino mass matrix:
$ZmZ^{\dagger} = $ diag. 
Substituting in (\ref{Gamma}),
we can compute the decay rates $\Gamma (h \rightarrow \chi^0 \nu)$
and $\Gamma (h \rightarrow \chi^+ \tau)$. Note that
by ``$h \rightarrow \chi^0 \nu$''  we mean 
$h \rightarrow \bar{\chi}^0 \nu$ and $h \rightarrow \chi^0 \bar{\nu}$.

 The $h_1 b \bar{b}$ coupling $\vec{\lambda} \cdot \hat{h}_1$
can be much larger in $R_p$ non-conserving theories than
in the MSSM. Decomposing $\vec{\lambda} =  
(\vec{\lambda} \cdot \vec{v}) \vec{v} /v^2 + 
(\vec{\lambda} \cdot \hat{\nu}) \hat{\nu}$,
(in the $\langle \tilde{\nu} \rangle = 0 $ basis this is 
$\vec{\lambda} = (h_b, \lambda')$), it follows that
\begin{eqnarray}
\vec{\lambda} \cdot \hat{h}_1 & =&  \frac{u_1}{det{\bf N}^{'}}
\left\{-\frac{g m_b}{\sqrt{2} m_W \cos \beta} 
\left[ (m_{A_1}^2 m_{A_2}^2 \sin \beta \cos \beta - m_{h_1}^2( S
 \tan \beta + m_Z^2
\cos \beta \sin \beta )) \right. \right. \nonumber \\
  & &\left. + m_Z^2 \sin \beta \cos \beta (m_{A_1}^2 +m_{A_2}^2 
- S /\cos^2 \beta) \right] \nonumber \\ 
& & \left.  +  \vec{\lambda} \cdot 
\hat{\nu} \ 
S \tan \beta (m_{h_1}^2 + m_Z^2(\sin^2 \beta
 - \cos^2 \beta))\sqrt{ \frac{ \delta_R}{1-\delta_R}} \right\} + loops
\end{eqnarray}
where $(\vec{\lambda} \cdot \hat{\nu})$
can be $\sim 1$ as discussed in section 3.
This expression simplifies  when
$R_p$ violation is small:
if $h_1 \rightarrow h$ when 
$ \delta_R  \rightarrow 0$ and  $ \vec{\lambda} \cdot \hat{\nu}
\rightarrow 0$,
 then $\vec{\lambda} \cdot \hat{h}_1  \rightarrow g 
m_b/(\sqrt{2} m_W \cos \beta)$
as expected in the MSSM.   If $h_3$ and $A_2$ are
the sneutrino components in the same limit, then
one can check that  $\vec{\lambda} \cdot \hat{h}_3  \rightarrow 0$.

\section{Results}
We express the masses and coupling constants
of the CP-even Higgses 
in terms of   tree-level input parameters  $m_{A_1}$, $m_{A_2}$, 
$\tan\beta$ and $\delta_R$.
When loop corrections are included there is an 
additional dependence on the soft
parameters $A$, $\mu_{I}$, $m_{Q}$, 
$m_{U}$ and $m_D$. This is the usual MSSM
set of input parameters, augmented by 
an additional CP-odd mass $m_{A_2}$,
and $\delta_R = $ the square of
an angle parametrising $R_p$ violation.
  We define $A_2$ to be the CP-odd scalar that
becomes the $\tilde{\nu}$ as $\delta_R \rightarrow 0$.
Which of the $A_i$ becomes the $\tilde{\nu}$
is important, because we expect the $R_p$ violating
effects to go to zero as $\msnu \rightarrow \infty$ for
all values of $\delta_R$.

It can be seen
from eqs. (\ref{mAmsnu})  and (\ref{vMv}) that requiring the invariant 
$S = \vec{v} \cdot $ [ {\bf M } $ ] \cdot \vec{v}/\chi^2 $  to be real,
dictates that the mass splitting  $|m_{A_1}^2 - m_{A_2}^2|$ and 
$\delta_{R}$  are not completely independent.
Fixing the value of this mass splitting will 
give a maximum allowed value of $\delta_R$ from requiring
that $S$ be real; also  
the mass splitting must be {\it greater} than a 
certain value  for a fixed value
of $\delta_R$.

\begin{figure}[h]

\vspace{-100pt}

\centerline{\hspace{-3.3mm}
\epsfxsize=8cm\epsfbox{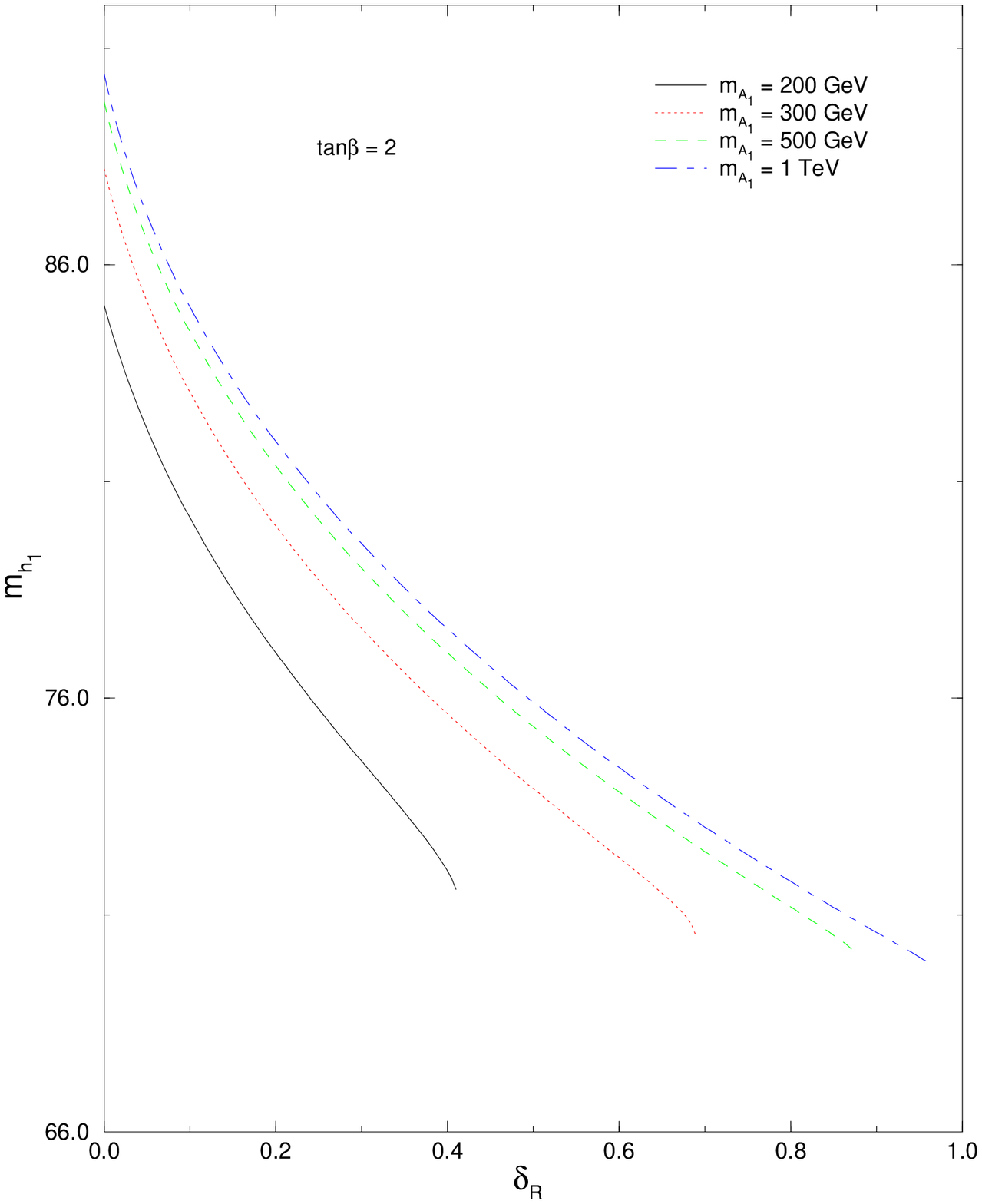}
\hspace{0.5cm}
\epsfxsize=8cm\epsfbox{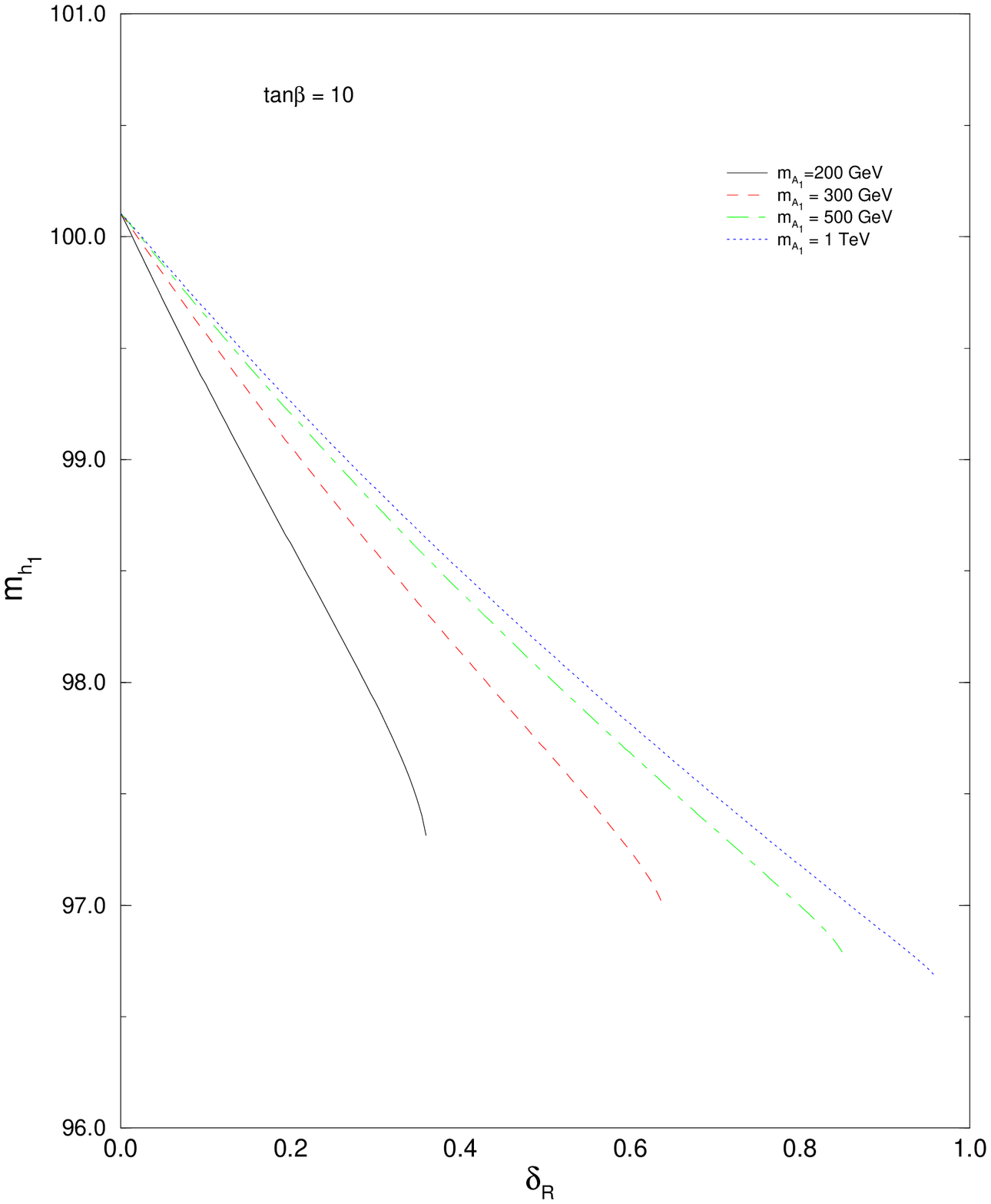}}
\vspace{0.5cm}
\centerline{\hspace{-3.3mm}
\epsfxsize=8cm\epsfbox{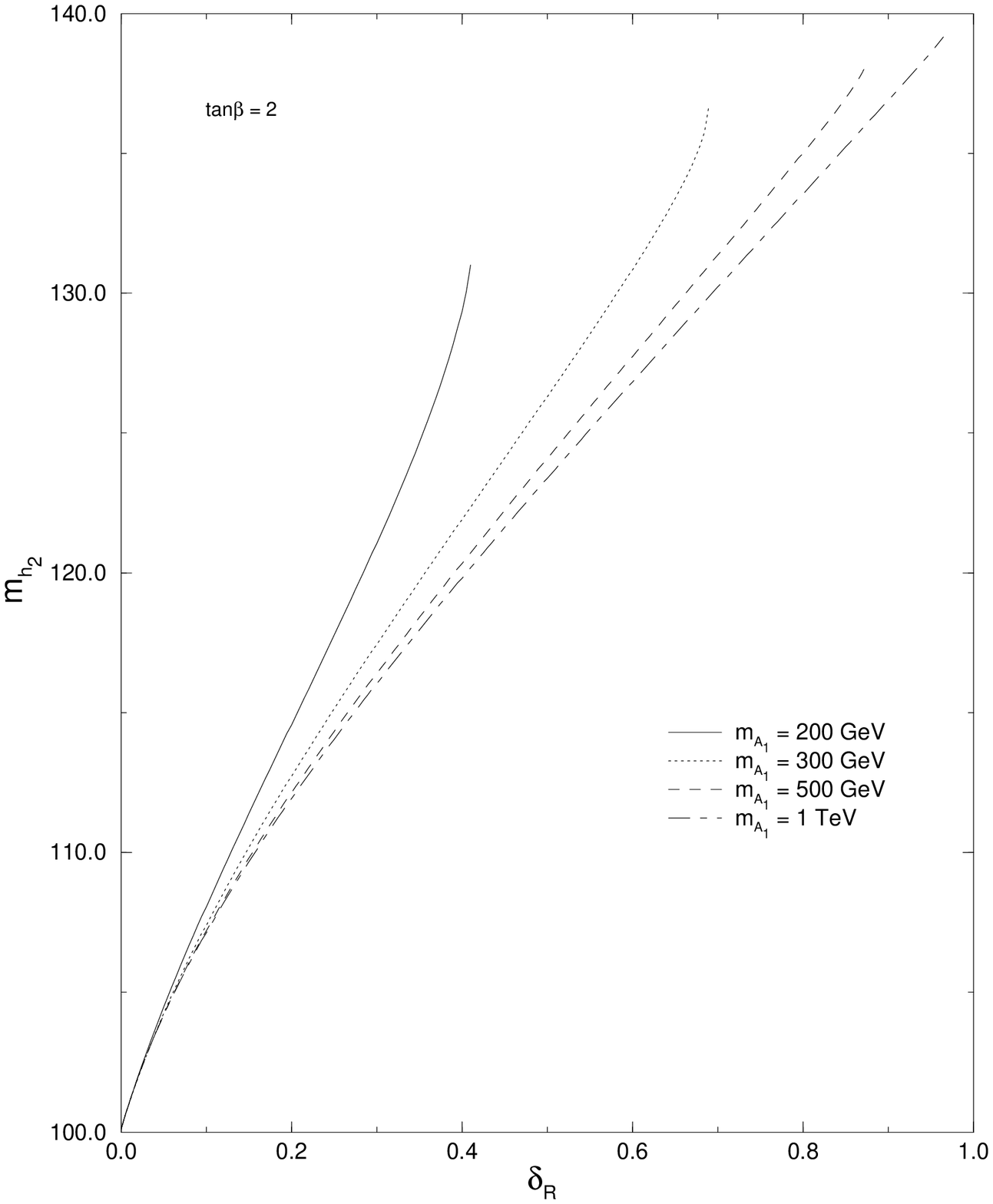}
\hspace{0.5cm}
\epsfxsize=8cm\epsfbox{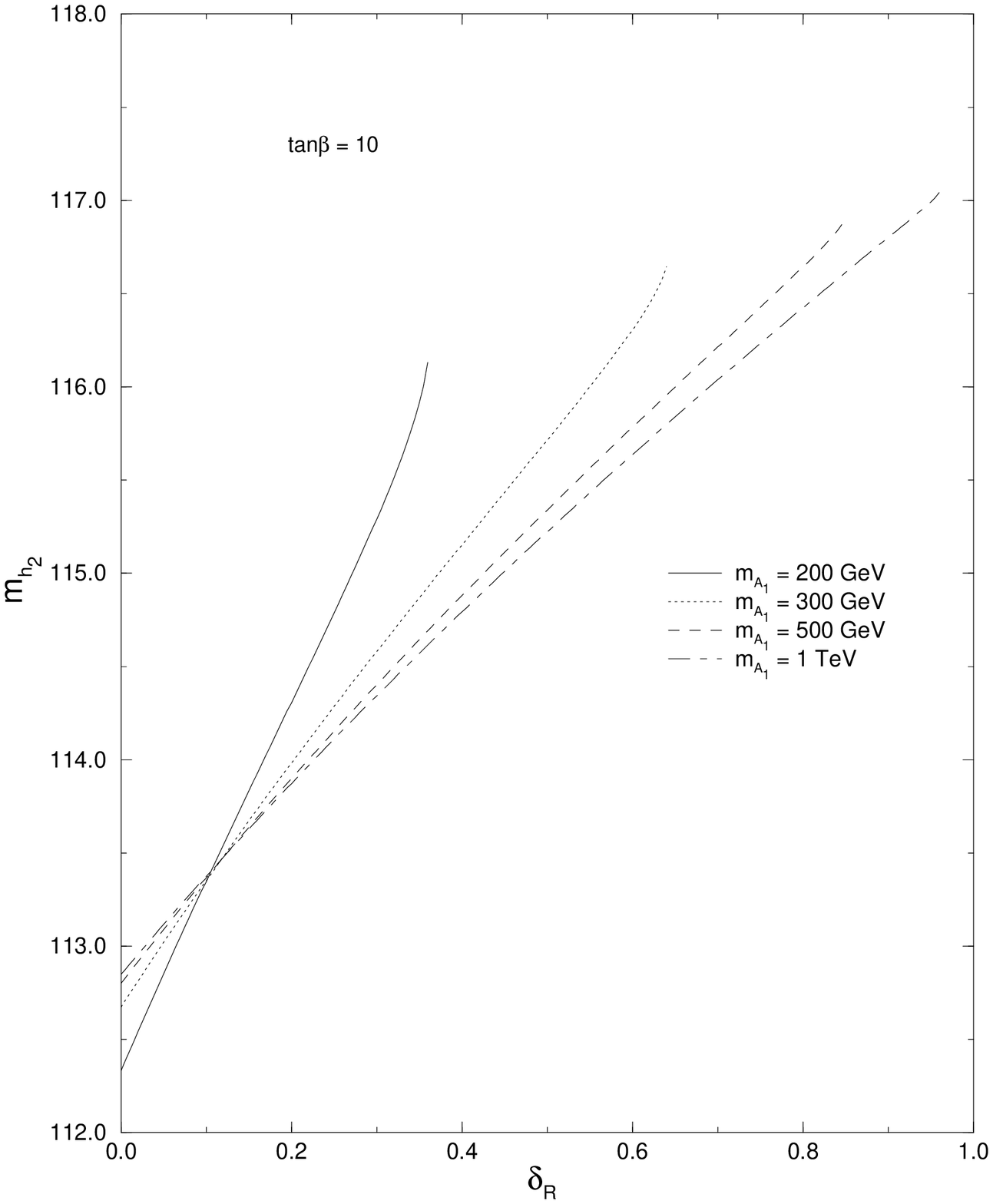}}

\caption{$m_{h_{1}}$ and $m_{h_{2}}$ as a function of $\delta_R$ for 
$m_{A_1}= 200,300,500,1000$ GeV and  $m_{A_2}=100$ GeV. The  input
parameters for the loop contributions to
the difference between the CP-even and CP-odd mass matrices
are $m_{Q}=500$ GeV, $m_U=m_D=300$ GeV, $A=200$ GeV, and
$\mu_{I} =(200,0)$. In the plots
on the left, $\tan\beta=2$; 
 $\tan\beta=10$ for the  plots on the right.}
\protect\label{mhiggs1delR}
\end{figure}

  In figure \ref{mhiggs1delR} we plot {\it $m_{h_1}$ and
 $m_{h_2}$ }  as a function of
$\delta_R$ for  values of $m_{A_1} =200, 300, 500, 1000$ GeV 
with  $\tan\beta=2, 10$ and $m_{A_2}=100$ GeV. 
%fixed
% and various choices of the other inputs.
We observe  the dependence of the maximum 
allowed value of $\delta_R$  on the mass splitting 
$|m_{A_1}^2 - m_{A_2}^2|$.
As the mass splitting increases the maximum 
value of $\delta_R$  also increases. Note that
$m_{h_1} (m_{h_2}) \rightarrow m_{A_2}$ in figure 3
as $\delta_R \rightarrow 0$  for $\tan\beta= 10 (2)$
because this is the Higgs which becomes the sneutrino
in this limit. 
We see that the lightest mass eigenvalue $m_{h_1}$, decreases
with $\delta_R$ for fixed CP-odd
masses. On the other hand, $m_{h_2}$ increases as a function of $\delta_R$. 
The effect of $\delta_R$ on
the heaviest eigenvalue $m_{h_{3}}$ is not very strong: 
we obtain $m_{h_{3}} \simeq m_{A_1}$ for all the allowed values of 
$\delta_R$.

% \begin{figure}[h]
% %\vspace{-80pt}
% \centerline{\mbox{\epsfig{file=mhiggs1delR2.eps,height=8cm,width=8cm}}
% \hspace{0.5cm}
% \mbox{\epsfig{file=mhiggs1delR10.eps,height=8cm,width=8cm}}}
% \vspace{1.5cm}
% \vspace{80pt}
% \centerline{\mbox{\epsfig{file=mhiggs2delR2.eps,height=8cm,width=8cm}}
% \hspace{0.5cm}
% \mbox{\epsfig{file=mhiggs2delR10.eps,height=8cm,width=8cm}}}
% %\vspace{-3cm}

\begin{figure}[h]

\vspace{-30pt}

\centerline{\hspace{-3.3mm}
\epsfxsize=8cm\epsfbox{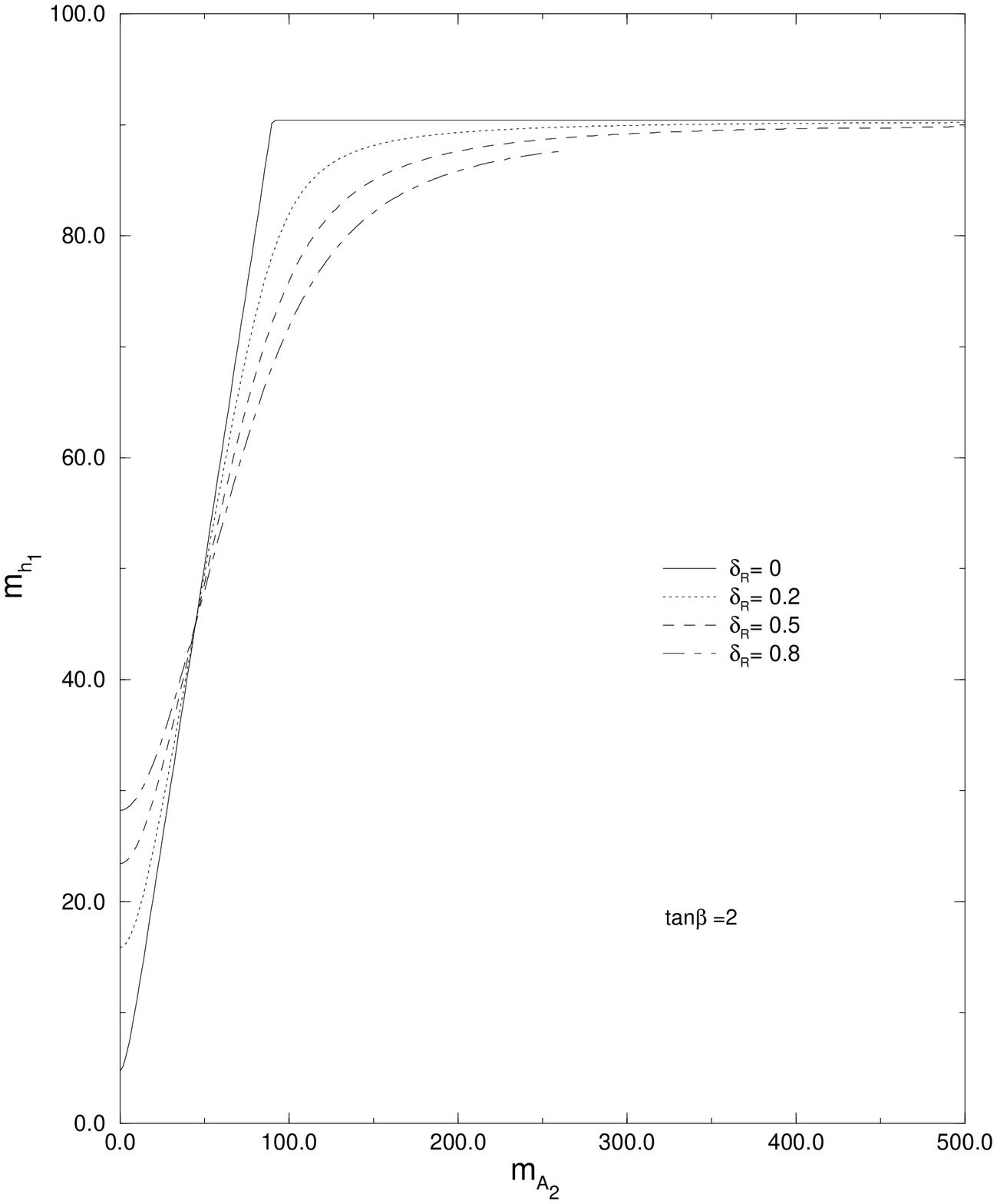}
\hspace{0.5cm}
\epsfxsize=8cm\epsfbox{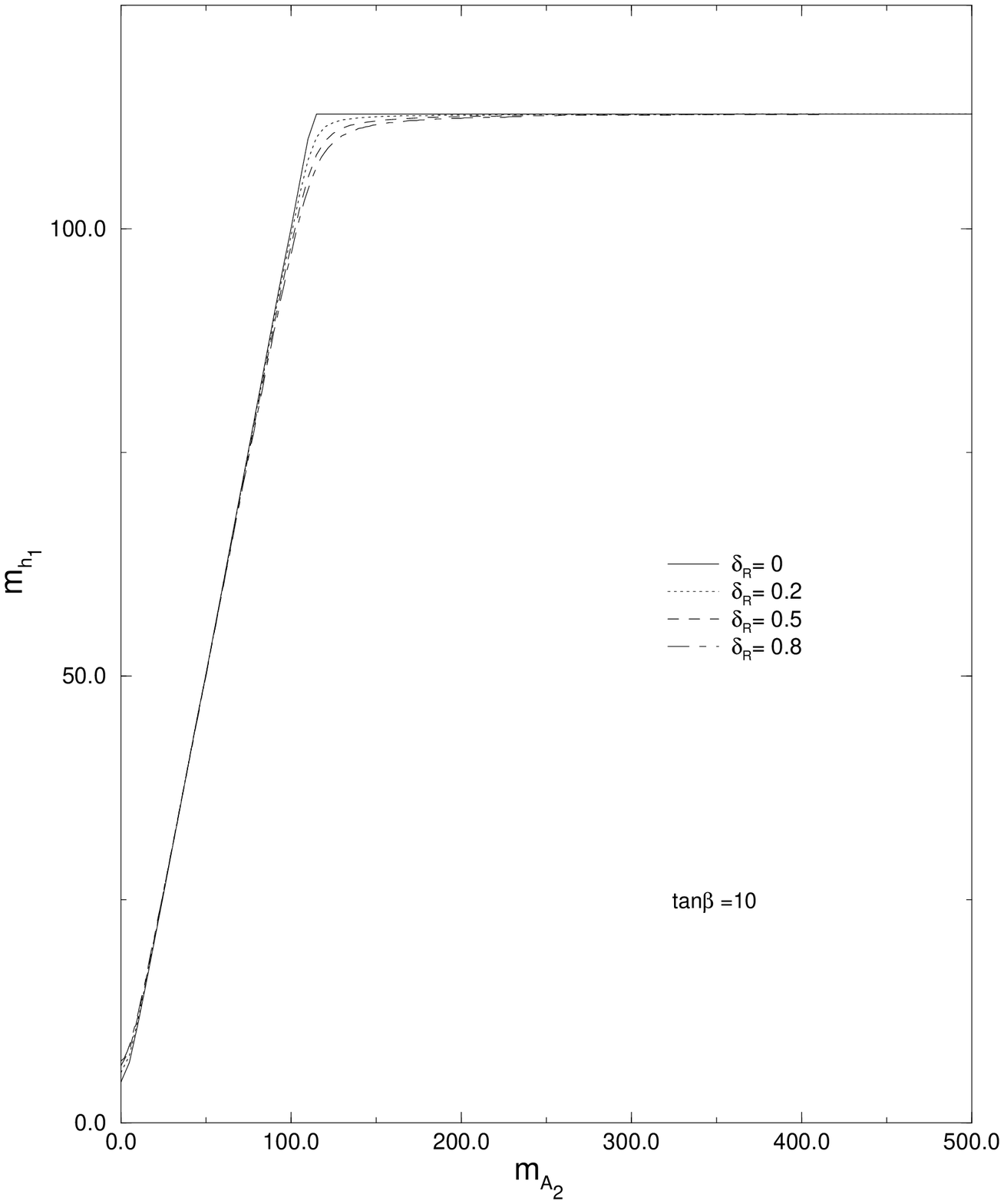}}

\caption{$m_{h_{1}}$ as a function of $m_{A_2}$ for 
$\delta_R =0,0.2,0.5,0.8$ and  $m_{A_1}=1$ TeV. The  input
parameters for the CP-even --- CP-odd
Higgs mass difference are $m_{Q}=500$ GeV, $m_U=m_D=300$ GeV, 
$A=200$ GeV, and $\mu_{I} =(200,0)$. 
 $\tan\beta=2$ on the left, and  $\tan\beta=10$ on the right.}
\protect\label{mhiggs1mA}
%\end{center}
\end{figure}

Conversely, in fig. \ref{mhiggs1mA} we present 
the variation of $m_{h_1}$ with respect to $m_{A_2}$ for 
$\delta_R =0,0.2,0.5,0.8$,  having fixed 
$m_{A_1}=1$ TeV, and for two values of 
$\tan\beta=2,10$. For each value of $\delta_R$ there is a
maximum allowed value of $m_{A_2}$. 
Recall that $A_2$ is a sneutrino component when
$\delta_R = 0$. As $m_{A_2} \rightarrow 0$, so does $m_{h_1}$ because
the lightest CP-even Higgs is the mode that becomes $\tilde{\nu}_R$
when $\delta_R = 0$. As $m_{A_2}$ increases,
the mode ``that would be the MSSM $h$ if $\delta_R = 0$''
becomes  the lightest Higgs and the plot flattens out. 
In the plot $m_{h_1}$ does not become exactly zero for $\delta_R=0$,
$m_{A_2}=0$ due to one-loop corrections 
proportional to $\lambda'^2$
from the squark-quark sector.
 
% \begin{figre}[h]
% \begin{center}
% \vspace{-30pt}

% \centerline{
% \mbox{\epsfig{file=pmhiggs1mA2.eps,height=8cm}}
% \hspace{0.5cm}
% \mbox{\epsfig{file=pmhiggs1mA10.eps,height=8cm}}}

As mentioned above, the loop corrections induced 
by values  $\lambda' \sim 1$
are similar to those induced by the top Yukawa 
for the Higgs which couples to the
up-type sector. The other  $R_p$-violating 
coupling we have introduced in the calculation
$\mu_1$ is always constrained by the neutrino 
mass to be sufficiently small that its 
contributions are negligible.

Including $R$-parity violation in the Higgs sector 
can be understood as having two effects on Higgs
production and decay.
It mixes the ``Higgses'' with the ``sneutrino'', and allows
new decay modes for the Higgs/sneutrino decay products.
There is  of course no distinction between a Higgs and a sneutrino 
in the presence of $R$-parity violating couplings;
by ``sneutrino'' we here mean the CP-even and -odd  mass eigenstates
that become the sneutrino in
the $\delta_R \rightarrow 0$ limit, and 
the ``Higgs'' is the CP-even mass eigenstate that
would be the Higgs in the same limit.
We define $A_2$ to be the CP-odd scalar that
becomes the $\tilde{\nu}_I$ as $\delta_R \rightarrow 0$.
Mixing the Higgs with the sneutrino means that
the  eigenstates $h_i$ can all be produced
 via $Z \rightarrow Zh_i$ and $Z \rightarrow h_i A_j$,
where $i: 1..3$ and $j:1..2$. 
All of the  $h_i$  
can decay to  $b \bar{b}$, and to $\chi^0  \nu$, 
 $\chi^+  \tau$ and $\chi \chi$  if these decay
modes are kinematically accessible.

The neutralino can decay in the detector,
via its production vertex (the neutralino
becomes a neutrino and an off-shell
Higgs, which can then decay to SM fermions).
So unless $\delta_R$ is uninterestingly small,
 the  $\chi^0 \nu$  and $\chi^0 \chi^0$ should be
visible. 

\begin{figure}[h]

\vspace{-30pt}

\centerline{\hspace{-3.3mm}
\epsfxsize=8cm\epsfbox{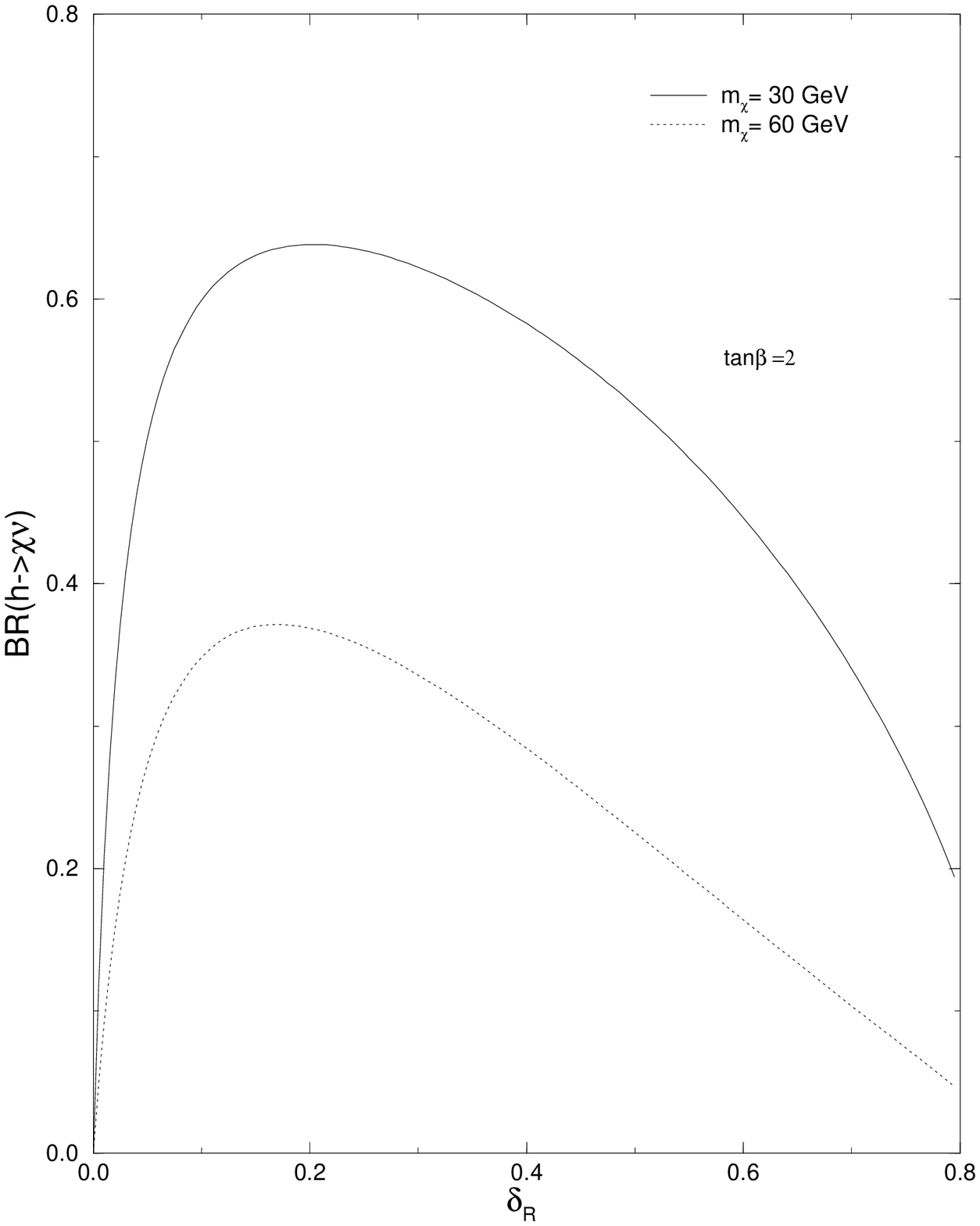}
\hspace{0.5cm}
\epsfxsize=8cm\epsfbox{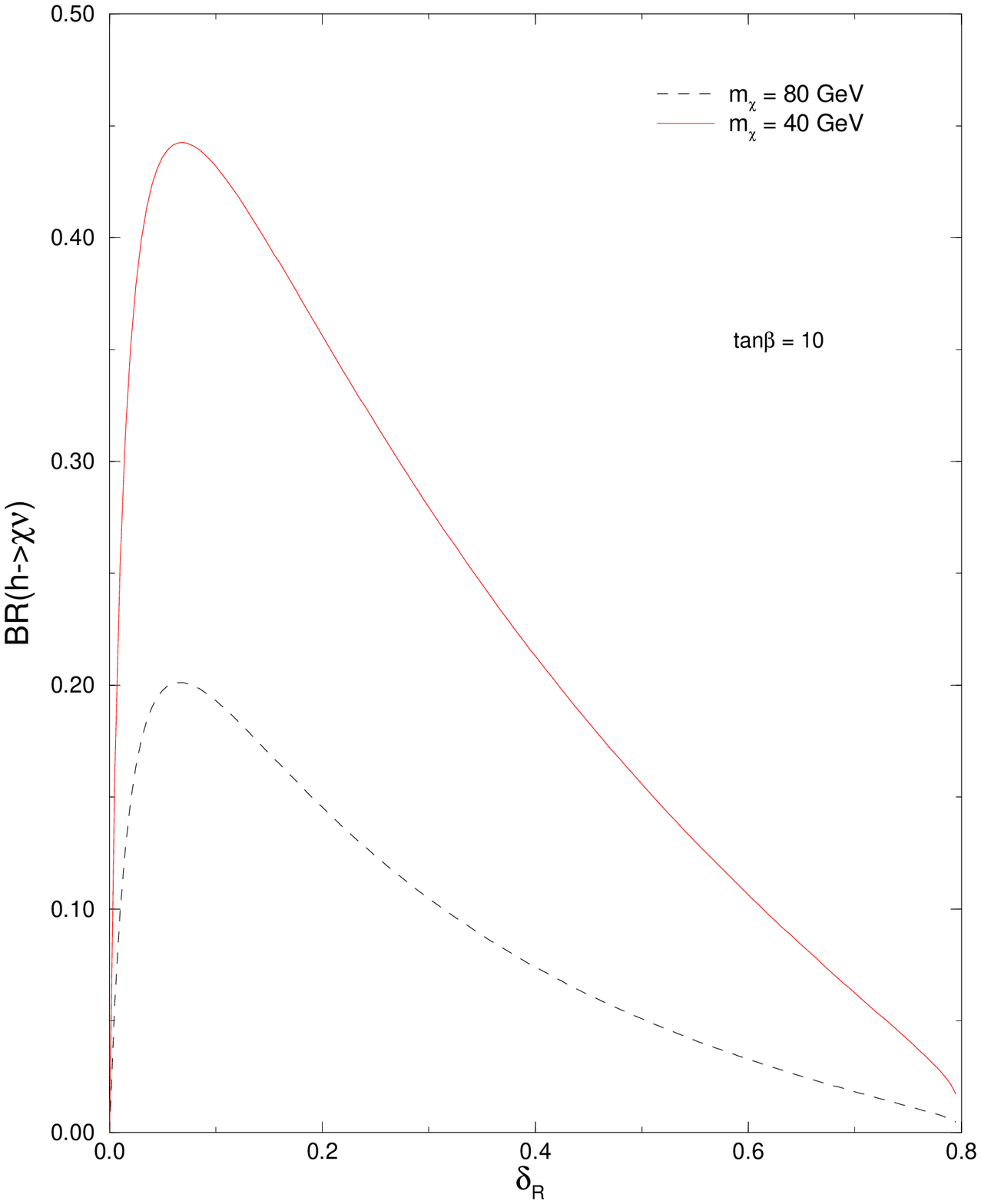}}

\caption{ The branching ratio for 
$h_1 \rightarrow \chi^0 \nu$ as a function of 
$\delta_R$, for different values of $m_{\chi}$ and
$\tan \beta$. 
We take $m_{A_1} = 500$ GeV, $m_{A_2} = 120$GeV ,
and the  input parameters 
for the loops contributing to the CP-even---CP-odd
Higgs mass difference are
 $A=0$, $\mu_I=(200,0)$,
$m_Q=500$ GeV, $m_U=m_D=300$GeV.
We take the total decay rate to be
$\Gamma_{tot}$ $ = \Gamma(h \rightarrow b \bar{b})$
$ + \Gamma(h \rightarrow  \chi^0 \nu)  $.}
\protect\label{figBRchinu}
\end{figure}

The new  $R_p$ violating 
decay modes  $h_1 \rightarrow \chi^+  \tau$
and  $h_1 \rightarrow \chi^0  \nu$ 
have been previously discussed  \cite{fer,drw}.
We plot the branching ratio to $\chi^0 \nu$
as a function of $\delta_R$ in figure (\ref{figBRchinu}).
We assume in this plot that the decays
to $\chi^+ \tau$ and $\chi^0 \chi^0$
are kinematically forbidden. 
As expected from equation (\ref{nudoth}),
the decay rate increases with $\delta_R$.
The  decrease at large $\delta_R$ is a consequence 
of our parametrisation. $\delta_R$ is $\sin^2$ of the angle
between the vectors $\vec{v}$ and $\vec{M}_u$;
as the angle increases to $\pi/2$ for fixed
$m_{A_1}$ and $m_{A_2}$,  $|\vec{M}_u|$ decreases.
So the $R_p$ violating mass term $|\vec{M}_u| \sqrt{\delta_R}$
decreases. For larger values of $\tan\beta$ the decay $h_1\rightarrow b \bar{b}$
is dominant.

Suppose now that the neutralino is also
heavier than $h_1$, so only the Standard Model
decays are available to the Higgs.
The $R_p$ violating  couplings
can still affect the production cross-section
of the Higgs, and therefore the
experimental lower limits on $m_h$.

The production cross-section for
$Z \rightarrow Z h$ can be parametrised
by $\xi^2 =  \sigma(Z \rightarrow Z h) /\sigma(Z \rightarrow Z h)_{SM}$.
See, $e.g.$, \cite{tampere} for experimental
limits on $\xi^2$.
In the MSSM, $\xi^2 = \sin^2 (\beta -\alpha)$. 
It can be very small in our $R_p$ violating
model  because it goes to zero as $\delta_R \rightarrow 0$
for  the CP-even Higgs  that becomes $\tilde{\nu}$ in
this limit.

\begin{figure}[h]
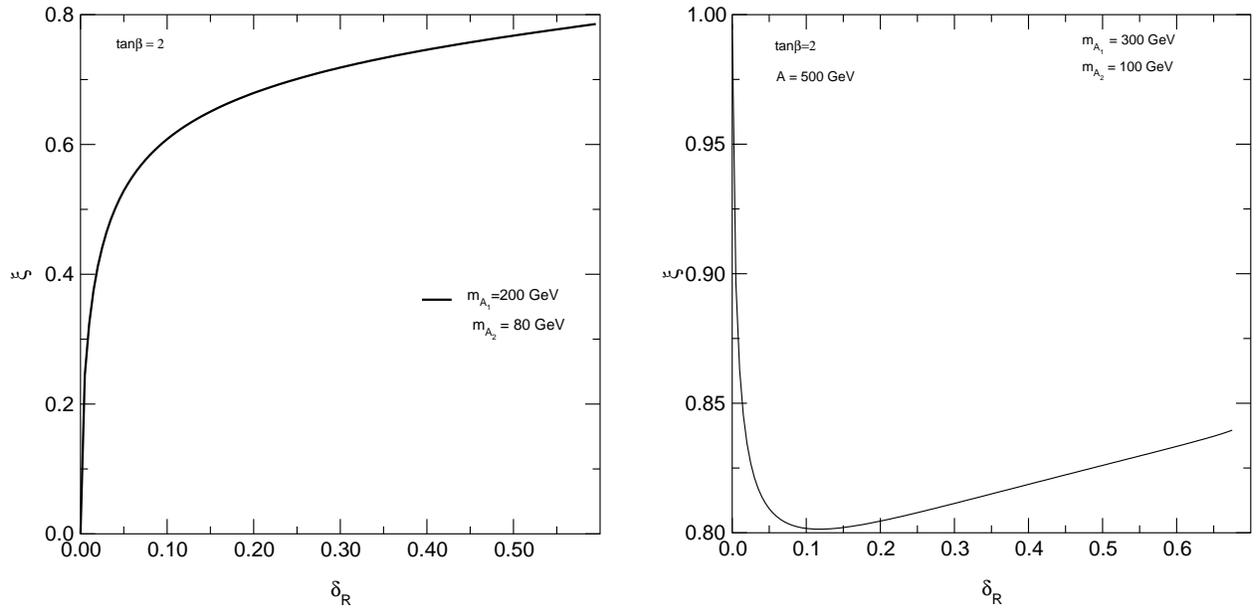

\begin{center}
\vspace{-30pt}

\centerline{
\mbox{\epsfig{file=sbmadelR22.eps,height=8cm}}
\hspace{0.5cm}
\mbox{\epsfig{file=sbmadelR23-A5.eps,height=8cm}}}

%\vspace{-3.5cm}

\caption{ The  ratio  $\xi$ as a function
of $\delta_R$, where $\xi = g_{ZZh}/g_{ZZh}^{SM}$
is the ratio of the $ZZh$ vertex  its
value in the SM. When the lightest CP-even
scalar $h_1$ corresponds to the sneutrino
in the $\delta_R \rightarrow 0$ limit,
$\xi$ can be small and 
goes to zero with $\delta_R$. If $m_{A_2} 
($which becomes $m_{\tilde{\nu}}$ when
$\delta_R = 0$) is large,
$\xi$ is near its MSSM value.  }

\protect\label{xi}
\end{center}
\end{figure}

If $m_{A_2}$ is heavier than the CP-even Higgs which 
becomes $h$ of the MSSM in the $\delta_R \rightarrow 0$ limit, 
then the CP-even Higgs corresponding to  $\tilde{\nu}_R$
in the same limit
will be heavier  than $m_{A_2}$ (see figure \ref{mhiggs1delR}).
In this case  the $Z \rightarrow Zh_1$ vertex
(equations \ref{RpZZh} and \ref{vdoth2}) does not differ much from its MSSM 
value. We plot $\xi  $  as a function of
$\delta_R$ on the RHS of figure \ref{xi} for
$m_{A_2} = 100$ GeV.
The present experimental lower limit on the Higgs
mass for $\xi   \sim .8$ is a few GeV below
the  $\xi = 1$ limit of 95.2 GeV \cite{tampere}.

Alternatively, if $m_{A_2}$ is light
\footnote{We assume that $\chi^0 \nu$  decays are 
nonetheless kinematically not allowed;
the stau is therefore the LSP, but  it decays so
is not cosmologically a problem.}, then  $\xi^2$ 
can be very small. For instance on the LHS of figure \ref{xi},
we plot $\xi$ for the  CP-even Higgs
which becomes part of the sneutrino when $\delta_R \rightarrow 0$.
As expected, $\xi$ is very small for
small $\delta_R$, because sneutrinos
in the MSSM are pair-produced.

Decreasing the $ZZh$ vertex  
would decrease the experimental Higgs mass
bound from this process; 
for $\xi \lsim .3$,  there is virtually no 
experimental lower limit \cite{tampere}. 
However, $\Gamma (Z \rightarrow h A)$ increases
as  $\Gamma (Z \rightarrow h Z)$ decreases,
so there should still be a bound on $m_h$.
The experimental
lower limit on $m_h$ from $Z \rightarrow h A$ 
is not trivial  to determine, because
the vertex and the two scalar
masses are independent  parameters. There are experimental
limits in the MSSM \cite{ZhA}, but
in this case the vertex and $m_h$ determine
$m_A$.  There are also bounds  on sneutrino
masses from $Z \rightarrow \tilde{\nu} \tilde{\nu}^*$ 
\cite{snusnu} in
models with trilinear  $R_p$
violation, but these assume that
the CP-even and CP-odd sneutrino  components
are degenerate. The experimental lower
limits on the Higgs masses in this model
are therefore unclear, but likely to be
lower than in the MSSM.

\section{Conclusions}

We have described the $R$-parity violation 
induced by the additional soft mass
terms in the scalar sector in terms of a basis-invariant 
quantity R (or $\delta_R$). This
eliminates the ambiguity usually present in these models 
when a specific Lagrangian basis for
the hypercharge $-1$ doublets is chosen. We have 
analysed the effects of the $R_p$-violating
couplings on the CP-even and CP-odd scalar masses 
to one-loop in a basis-invariant way.
We have also calculated the $R_p$ conserving
and $R_p$ violating branching ratios of the lightest Higgs boson
as a function of the basis-invariant quantity $\delta_R$. 
We have identified the
 regions of parameter space for which
the  decay  modes of the Higgs boson are not those of 
 the Standard Model Higgs.
We have also calculated the production cross section  
as a function of $\delta_R$,  and 
found that this can be strongly modified with 
respect to the $R_p$ conserving case
when the lightest Higgs boson is mostly ``sneutrino-like''.
The LEP lower bound on the Higgs mass in this
model can therefore be lower than in the MSSM.  
 
\section*{Acknowledgements}
S.D. and M.L. would like to thank J.R. Espinosa, P. Gambino, 
G. Ganis, G. Giudice and C. Wagner for useful discussions.  
N.R. wants to thank the CERN TH Division for kind hospitality.
This work was supported in part by DGESIC under Grant No. 
PB97-1261, by DGICYT under contract PB95-1077 and by EEC under the 
TMR contract ERBFMRX-CT96-0090.

\section*{Appendix A: $\langle \tilde{\nu} \rangle \neq 0$ basis}

We take the potential $V$ to be the sum of equations
(\ref{Vtree}) and (\ref{Vloop}). 
We define $v  = \sqrt{v_d^2 + v_L^2}$ $ = \sqrt{v_I v^I}$, and
\beq
\tan \beta = \frac \chi v .
\eeq
 We write the stop and sbottom masses as
\beq
 m_{\tilde{1},\tilde{2}}^2 = \frac{1}{2} \left\{ M_L^2 + M_R^2 \pm
 \sqrt{ (M_L^2 - M_R^2)^2 + 2X^2} \right\}
\eeq
where for the stops 
\beq
M_L^2 = m_Q^2 + h_t^2 \chi^2/2 + (g^2 - {1\over 3}g'^2)(v^2 -\chi^2)/8
\eeq
\beq
M_R^2 = m_U^2 + h_t^2 \chi^2/2 + g'^2(v^2-\chi^2)/6
\eeq
 and 
\beq
X_t^2 = (A_t \chi + h_t \mu \cdot v)^2
\eeq

 For the sbottoms
 \beq
M_L^2 = m_Q^2 + (\lambda \cdot v)^2/2 - (g^2 + {1\over 3}g'^2)(v^2 -\chi^2)/8
\eeq
\beq
M_R^2 = m_D^2 + (\lambda \cdot v)^2/2 -  g'^2(v^2-\chi^2)/12
\eeq
 and 
\beq
X_b^2 = (A_b \cdot v + \lambda \cdot  \mu  \chi)^2~~~.
\eeq

The one loop minimization conditions can be expressed in terms of
the CP odd mass matrix $M_{ij}$ as in equations (\ref{mincdn1})
and (\ref{mincdn2}):
\beq
 M_{uu} +  \frac{M_{uI}v^I}{\chi} = 0 
\label{Amincdn1}
\eeq
\beq
 M_{uI} +  \frac{{\bf M}_{IJ}v^J}{\chi} = 0 
\label{Amincdn2}
\eeq

The CP-odd mass matrix is:

\beq
\left[
\begin{array}{cc}
M_{uu} & \left( \begin{array}{cc} M_{ud} & M_{uL} 
           \end{array} \right) \\
 \left( \begin{array}{c} M_{ud} \\ M_{uL}  \end{array} \right) & 
            \left( \begin{array}{cc} {\bf M}_{dd} & {\bf M}_{dL} \\ 
               {\bf M}_{dL} & {\bf M}_{LL} \end{array} \right)
\end{array}
\right]
\eeq
where the components are 
\bea
M_{uu} &=& m_u^2 -\frac{ m_Z^2 \cos 2\beta}{2} 
%- \cot \beta (m_3^2)
+ \left\{ \frac{3 h_t^2 }{32 \pi^2}  
[f(m_{\tilde{t}_1}) + f(m_{\tilde{t}_2}) - 2 f(m_t)] 
+ (\lambda^I\mu_I)^2  D_b   +  A_t^2 D_t \right\} 
\\
M_{uI} &=& B_I 
+ \left\{ ( h_t A_t  D_t)  \mu_I
+( \mu^J\lambda_J D_b)  A^b_I\right\} \\
{\bf M}_{IJ} &=&   [m^2_L]_{IJ} + \frac{ m_Z^2 \cos 2\beta}{2}  \; \delta_{IJ}
 \nonumber \\
& & ~~~~~+ \left\{ \frac{3  }{32 \pi^2}  
[f(m_{\tilde{b}_1}) + f(m_{\tilde{b}_2}) - 2 f(m_b)] \lambda_I
\lambda_J 
+ h_t^2  D_t \mu_I \mu_J   +   D_b A^b_I A^b_J \right\} \ .
\eea

We have defined
\beq
f(m) = 2 m^2 \left(  \log \frac{m^2}{Q^2} -1 \right) \ , 
\eeq
where $Q^2$ is the renormalisation scale in the $\overline{MS}$ scheme, 
and 
\bea
D_t &\equiv& \frac{3}{32 \pi^2} \frac{1}
{ \Delta_t} [f(m_{\tilde{t}_1}) - f(m_{\tilde{t}_2})]  \ ,\\
D_b &\equiv& \frac{3}{32 \pi^2} \frac{1}
{\Delta_b} [f(m_{\tilde{b}_1}) - f(m_{\tilde{b}_2})]  
\eea
with 
\beq
\Delta_t = m^2_{\tilde{t}_1} - m^2_{\tilde{t}_2}  \ , 
\Delta_b = m^2_{\tilde{b}_1} - m^2_{\tilde{b}_2} \ .
\eeq

The CP-even scalar mass matrix is:

\bea 
M^{'}_{uu} &=& M_{uu} +  m_Z^2 \sin^2 \beta +
 \left\{  \frac{3 h_t^4}{16\pi^2}  \chi^2 \log\frac{\mst^2
      \msT^2}{m_t^4} \right. \nonumber \\  & & \left.+
 \frac{3 h_t^2}{16\pi^2}  \frac{2 A_t X_t \chi}{\Delta_t}
    \log\frac{\mst^2}{\msT^2} 
 + A_t^2 X^2_t g(\mst,\msT)
 +  (\lambda_I \mu^I)^2 X^2_b
 g(\msb,\msB) \right\}   \\  %\nonumber
%[m_H^2]_{uI}
M^{'}_{uJ} &=& M_{uJ} - m_Z^2 \cos \beta \sin \beta \frac{v_J}{v}
+  \left\{ \left[ \frac{3}{16\pi^2} ( v_K \lambda^K) ( \mu_K \lambda^K)
  \frac{ X_b }{\Delta_b}
    \log\frac{\msb^2}{\msB^2} \right] \lambda_J  
\right. \nonumber  \\  %\nonumber
 & & \left. + \frac{3}{16\pi^2} \left[ \frac{ \chi h_t^3 X_t }{\Delta_t}
    \log\frac{\mst^2}{\msT^2} \right] \mu_J 
+ \left[  X^2_b (\lambda_K \mu^K) g(\msb,\msB) \right]  A^b_J
 + \left[  X^2_t h_t A_t  g(\mst,\msT) \right]  \mu_J
 \right\} \\ %\nonumber
% [m_H^2]_{IJ}
{\bf M}^{'}_{IJ} &=&   {\bf M}_{IJ} +   m_Z^2 \cos^2 \beta \frac{v_I v_J}{v^2}
 + \left\{ \left[ \frac{3}{16 \pi^2}  (v_K \lambda^K)^2 \log\frac{\msb^2
      \msB^2}{m_b^4} \right] \lambda_I \lambda_J \right.  \nonumber \\ 
  & & \left. + 
\left[ \frac{3}{16 \pi^2} ( v_K \lambda^K)  \frac{ X_b }{\Delta_b}
    \log\frac{\msb^2}{\msB^2}\right] (\lambda_I A^b_J + \lambda_J A^b_I)  
 + \left[  X^2_b g(\msb,\msB) \right] A^b_I A^b_J \right.   \nonumber \\
 & & + \left. \left[ X^2_t h_t^2  g(\mst,\msT) \right] 
\mu_I \mu_J \right\} 
\eea

where
\beq
g(m_1,m_2) =  \frac {3}{16 \pi^2} \frac{1}{(m_1^2-m_2^2)^2}
\left[2 - \frac{m^2_1 + m^2_2}
{m^2_1 - m^2_2} \log \frac{m_1^2}{m_2^2} \right]  \ .
\eeq

\section*{Appendix B: $\langle \tilde{\nu} \rangle = 0$}

This Appendix contains one-loop formulae for the minimisation
conditions, the CP-odd mass matrix  and 
the CP-even mass matrix, in the basis where the
sneutrino vev $\langle \tilde{\nu} \rangle = v_L$
is zero at one loop.

We define  the up-type Higgs vev to be $\chi/\sqrt{2}$,
and the down-type vev to be $v/\sqrt{2}$, so
\beq
\tan \beta = \frac \chi v
\eeq
and $X_t = A_t \chi + h_t \mu v$.

In this basis, we can safely neglect the loop corrections due to 
$h_b$ and the soft trilinear coupling $A^b_d \propto h_b$, 
since they are constrain to be small by the $b$-quark 
mass $m_b=-h_b v/\sqrt{2} \ll v$.
If $m_1^2 \equiv [{\bf m}_L^2]_{dd}, m_2^2 \equiv m_u^2$ and
$m_3^2 \equiv B_d$ are tree-level Higgs mass
terms, $\mu \equiv  \mu_0$ ($\epsilon \equiv \mu_1$) is the $R_p$ 
conserving (violating)
superpotential mass, and $A' \equiv A^b_L$, then
the minimisation  conditions are

\bea
m_1^2 &=& - m_3^2 \tan \beta - \frac{1}{2} m_Z^2 \cos 2\beta 
+ \delta m_1^2 \\
m_2^2 &=&  - m_3^2 \cot \beta + \frac 1 2 m_Z^2 \cos 2\beta 
 + \delta m_2^2 \\
{[{\bf m}_L^{2}]}_{dL}
 &=& - B_L  \tan \beta - h_t \epsilon D_t
\frac{X_t}{v} - A' \lambda' \epsilon \tan \beta D_b
\eea
where
\bea
\delta m_1^2 &=& - h_t \mu D_t \frac{X_t}{v} \ , 
\\
\delta m_2^2 &=& - \frac{3}{32 \pi^2} h_t^2 
[f(m_{\tilde{t}_1}) + f(m_{\tilde{t}_2}) - 2 f(m_t)] 
-  A_t D_t \frac{X_t}{\chi}
- \lambda'^2 \epsilon^2  D_b  \ .
\eea

We have defined $D_t, D_b$ and $f(m)$ 
as in the previous appendix.

The 
CP-odd scalar mass matrix elements are

\bea
M_{uu} &=& - \cot \beta (m_3^2  + h_t \mu A_t D_t) 
\\
M_{ud} &=& m_3^2 + h_t \mu A_t D_t \\
M_{uL} &=& B_L + h_t A_t \epsilon D_t 
+ A' \epsilon \lambda' D_b \\
{\bf M}_{dd} &=& - \tan \beta (m_3^2 + h_t \mu A_t D_t) \\
{\bf M}_{dL} &=& [{\bf m}_L^2]_{dL}  + h_t^2 \mu \epsilon D_t \\
{\bf M}_{LL} &=& m_{A_1}^2 + m_{A_2}^2 + (m_3^2 + h_t \mu A_t D_t)
\frac{2}{\sin 2\beta}
\eea
with 
\bea
m_3^2 & =& - \frac 1 2 \left\{ (m_{A_1}^2 + m_{A_2}^2) \sin \beta \cos \beta +
 2 h_t \mu A_t D_t  \right\} \\ 
& & - \frac{\sin \beta \cos \beta}{ 2} 
\sqrt{(m_{A_1}^2 - m_{A_2}^2)^2  - \frac{4}{ \cos^2 \beta} 
(B_L + 
h_t A_t \epsilon D_t + \lambda' A' \epsilon D_b)^2}
\nonumber 
\eea

The 
CP-even scalar mass matrix $M'$ is:

\bea 
M'_{uu} &=& - \cot \beta (m_3^2 + h_t \mu A_t D_t) + m_Z^2 \sin^2 \beta
+ \lambda'^4 \epsilon^4 \chi^2 g(\msb,\msB)
\nonumber \\ 
&+&
 \frac{3}{16\pi^2} \left[h_t^4 \chi^2 \log\frac{\mst^2
      \msT^2}{m_t^4} + \frac{2 h_t^2 A_t X_t \chi}{\Delta_t}
    \log\frac{\mst^2}{\msT^2} \right] + A_t^2 X^2_t g(\mst,\msT)  
\\
M'_{ud} &=& m_3^2 - \frac{m_Z^2}{2} \sin 2\beta + h_t \mu A_t D_t +
  \frac{3 h_t^3 \mu X_t}{16\pi^2 \Delta_t} \chi
  \log\frac{\mst^2}{\msT^2} + h_t \mu A_t X_t^2 g(\mst,\msT) 
\\
M'_{uL} &=&  B_L  + h_t A_t \epsilon D_t 
+ \frac{3}{16\pi^2} h_t^3  \frac{\epsilon X_t \chi}{\Delta_t} 
\log\frac{\mst^2}{\msT^2}
+ h_t A_t \epsilon X_t^2 g(\mst,\msT)
\nonumber \\ 
& & + A' \epsilon \lambda' D_b 
+ A'  \epsilon^3 \lambda'^3 \chi^2 g(\msb,\msB) 
\\
{\bf M}'_{dd} &=& - \tan \beta (m_3^2 + h_t \mu A_t D_t) + m_Z^2 \cos^2
\beta + h_t^2 \mu^2 X_t^2 g(\mst,\msT)
\\
{\bf M}'_{dL} &=& m_{dL}^2 + h_t^2 \mu \epsilon D_t
+ h_t^2 X_t^2 \mu \epsilon g(\mst,\msT) \\
{\bf M}'_{LL} &=& {\bf M}_{LL} + h_t^2 \epsilon^2 X_t^2  g(\mst,\msT)
+ A'^2  \epsilon^2 \lambda'^2 \chi^2 g(\msb,\msB)
\eea
where $\Delta_t$ and  
$g(m_1,m_2)$ are as defined in the previous appendix.

\section*{Appendix C: Some equations}

The normalisation factors for the basis-independent
Higgs mixing angles, at tree level,  are
\beq
\frac{u}{det[ {\bf N'}]} =  -\frac{1}{\sqrt{(det \; {\bf N'})^2 + V^{'2}}};
\eeq
where {\bf N}$^{'} = m_h^2$ {\bf I} - {\bf M}$^{'}$, and
\beq
\vec{V}' =  {\bf N'}   \cdot \varepsilon \cdot  \vec{M}^{'}_{u}
\label{V'}
\eeq
For $h = h_1, h_2$ or $h_3$, 
\beq
\begin{array}{lcl}
det \;{\bf N'} & =& m_h^4 - m_h^2 (m_Z^2 \cos^2\beta + m_{A_1}^2 + m_{A_2}^2) 
+ \sin^2\beta m_{A_1}^2 m_{A_2}^2  + S (m_h^2 - m_Z^2) \nonumber \\ & & 
+ (m_{A_1}^2+ m_{A_2}^2) m_Z^2 \cos^2\beta;
\end{array}
\eeq
and
\beq
\begin{array}{lcl}
V^{'2} &  =&  m_h^4 (S^2 \tan^2\beta/(1 - \delta_R)+ 2 m_Z^2 S \sin^2\beta  
+ m_Z^4 \cos^2 \beta \sin^2 \beta ) \\ & &
+m_{A_1}^4 m_{A_2}^4 \sin^2\beta \cos^2 \beta + 2m_Z^2\cos^2\beta \sin^2\beta 
(m_{A_1}^2 m_{A_2}^2 - m_h^2 m_Z^2 )(m_{A_1}^2 + m_{A_2}^2 - S/\cos^2\beta)  
\nonumber \\ & &
+ m_Z^4 \cos^2\beta \sin^2 \beta [(m_{A_1}^2 + m_{A_2}^2 - S)^2 
-2 m_{A_1}^2 m_{A_2}^2 \sin^2\beta  - S^2 \tan^4 \beta /(1-\delta_R)]+  
\nonumber \\ & & 
m_Z^4 S^2 \cos^2\beta  \sin^2\beta  \delta_R /(1- \delta_R) 
- 2m_Z^2 \sin^2\beta m_h^2 S^2 \delta_R/(1 - \delta_R)   
\nonumber \\ & & 
- 2 m_h^2 m_{A_1}^2 m_{A_2}^2 \sin^2\beta (S + 2 m_Z^2 \cos^2\beta) 
-2 S^2 m_Z^4 \sin^4\beta  \delta_R/(1-\delta_R)
\end{array}
\eeq

For the CP-odd Higgses, the normalisation factor is
\beq
n = \frac{-1}{\sqrt{(det{\bf N})^2 + V^2}}
\eeq
where {\bf N} = $m_a^2$ {\bf I} - {\bf M}, $\vec{V} =  $
{\bf N} $ \cdot \varepsilon \cdot \vec{M}_u$, $a = $ either $A_1 $ or $ A_2$,
  and
\beq
det \;{\bf N} =  m_a^4 - m_a^2 (m_{A_1}^2 + m_{A_2}^2 - S) + 
m_{A_1}^2 m_{A_2}^2 \sin^2 \beta
\eeq
and 
\beq
V^2 = m_a^4 S^2 \tan^2\beta /(1-\delta_R) - 2 m_a^2 m_{A_1}^2 m_{A_2}^2 S 
\sin^2 \beta 
+ m_{A_1}^4 m_{A_2}^4 \sin^2 \beta  \cos^2\beta
\eeq

\end{document}